\documentclass[sigconf]{acmart}
\usepackage{booktabs} 
\usepackage{tikz}
\usetikzlibrary{bayesnet}
\usepackage{algorithm}
\usepackage{multirow}
\usepackage{amsmath}
\usepackage[noend]{algpseudocode}
\usepackage{subfigure}
\usepackage{cleveref}
\usepackage{dblfloatfix}

\setcopyright{rightsretained}

\acmDOI{10.475/123_4}

\acmISBN{123-4567-24-567/08/06}

\acmYear{2017}
\copyrightyear{2016}
\acmPrice{15.00}
\renewcommand\footnotetextcopyrightpermission[1]{} 
\begin{document}
\title{Improving Latent User Models in Online Social Media}

\author{Adit Krishnan}
\affiliation{%
 \institution{Dept of Computer Science, University of Illinois at Urbana-Champaign}
}
\email{aditk2@illinois.edu}
\author{Ashish Sharma}
\affiliation{%
 \institution{Dept of Computer Science, Indian Institute of Technology Kharagpur}
}
\email{ashish4@illinois.edu}
\author{Hari Sundaram}
\affiliation{%
 \institution{Dept of Computer Science, University of Illinois at Urbana-Champaign}
}
\email{hs1@illinois.edu}
%
%
%
%

\renewcommand{\shortauthors}{Author 1 et al.}

\begin{abstract}
Modern social platforms are characterized by the presence of rich user-behavior data associated with the publication, sharing and consumption of textual content. Users interact with content and with each other in a complex and dynamic social environment while simultaneously evolving over time. In order to effectively characterize users and predict their future behavior in such a setting, it is necessary to overcome several challenges. Content heterogeneity and temporal inconsistency of behavior data result in severe sparsity at the user level. In this paper, we propose a novel mutual-enhancement framework to simultaneously partition and learn latent activity profiles of users. We propose a flexible user partitioning approach to effectively discover rare behaviors and tackle user-level sparsity.

We extensively evaluate the proposed framework on massive datasets from real-world platforms including Q\&A networks and interactive online courses (MOOCs). Our results indicate significant gains over state-of-the-art behavior models (~15\% avg ) in a varied range of tasks and our gains are further magnified for users with limited interaction data. The proposed algorithms are amenable to parallelization, scale linearly in the size of datasets, and provide flexibility to model diverse facets of user behavior. An updated, published version of this paper can be found here \cite{cikm2018}.
\end{abstract}

%
%
%

\vspace{-4pt}
\maketitle


\section{Introduction}
\label{sec:introduction}


This paper addresses the problem of developing robust statistical representations of participant behavior and engagement in online knowledge-exchange networks. Examples of knowledge-exchange networks include interactive MOOCs (Massive Online Open Courses), where participants interact with lecture content and peers via course forums, and community Q \& A platforms such as Stack-Exchanges\footnote{\url{https://stackexchange.com}}.
We would like statistical representations to provide insight into dominant behavior distributions, and understand how an individual's behavior evolves over time. Understanding and profiling time-evolving participant behavior is important in several applications; for instance, pro-actively identifying unsatisfactory student progress in MOOCs may lead to redesign of the online learning experience. The dynamic nature and diversity of user activity in such learning environments pose several challenges to profiling and predicting behavior. 
\begin{figure}[t]
  \subfigure[b][]{
   \centering
   \includegraphics[width=0.462\linewidth, trim={1.5cm 1.50cm 1.52cm 1.3cm},clip]{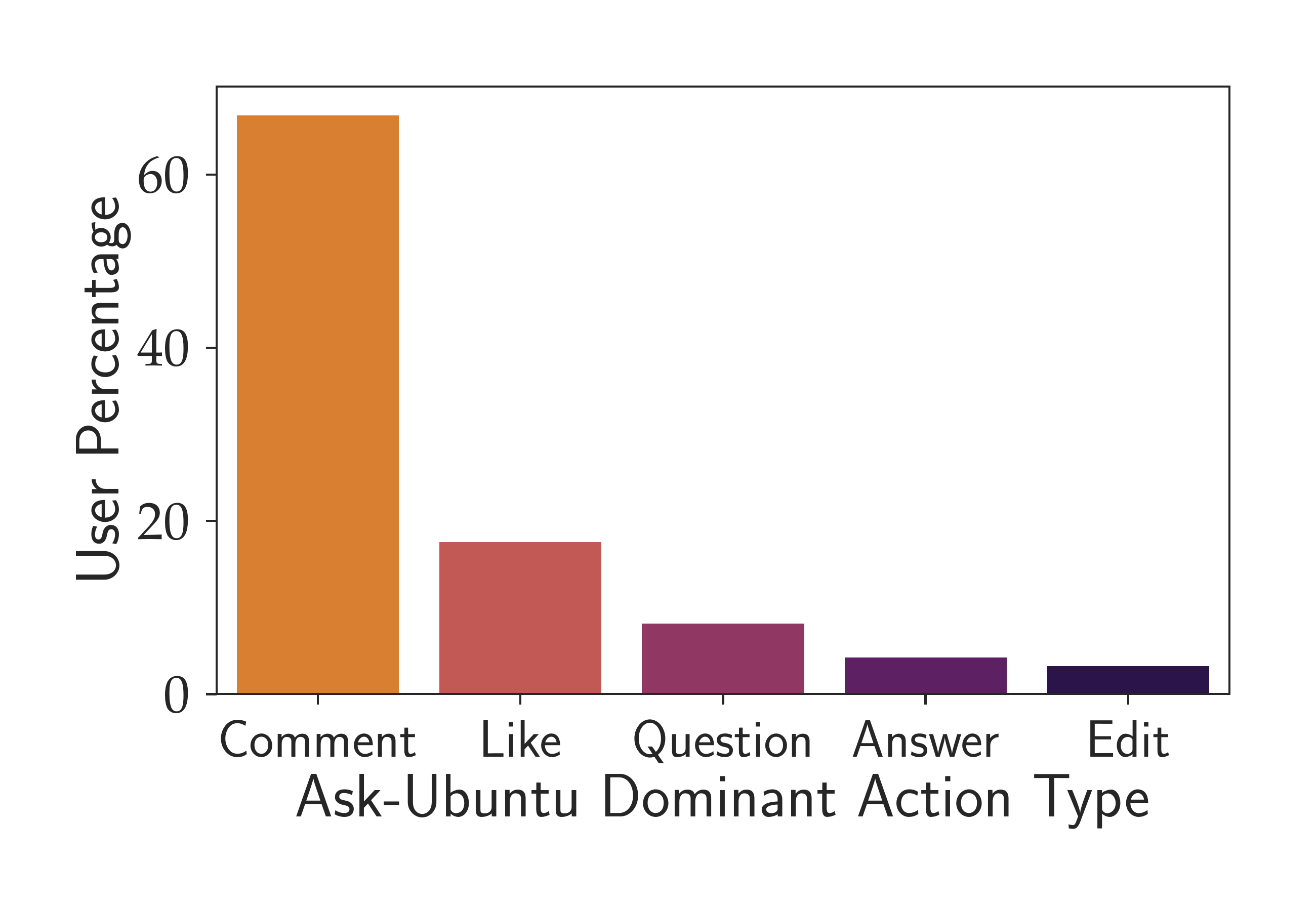}
	 \label{skewa}
  }
  \subfigure[b][]{
   \centering
   \includegraphics[width=0.462\linewidth, trim={1.5cm 1.50cm 1.52cm 1.3cm},clip]{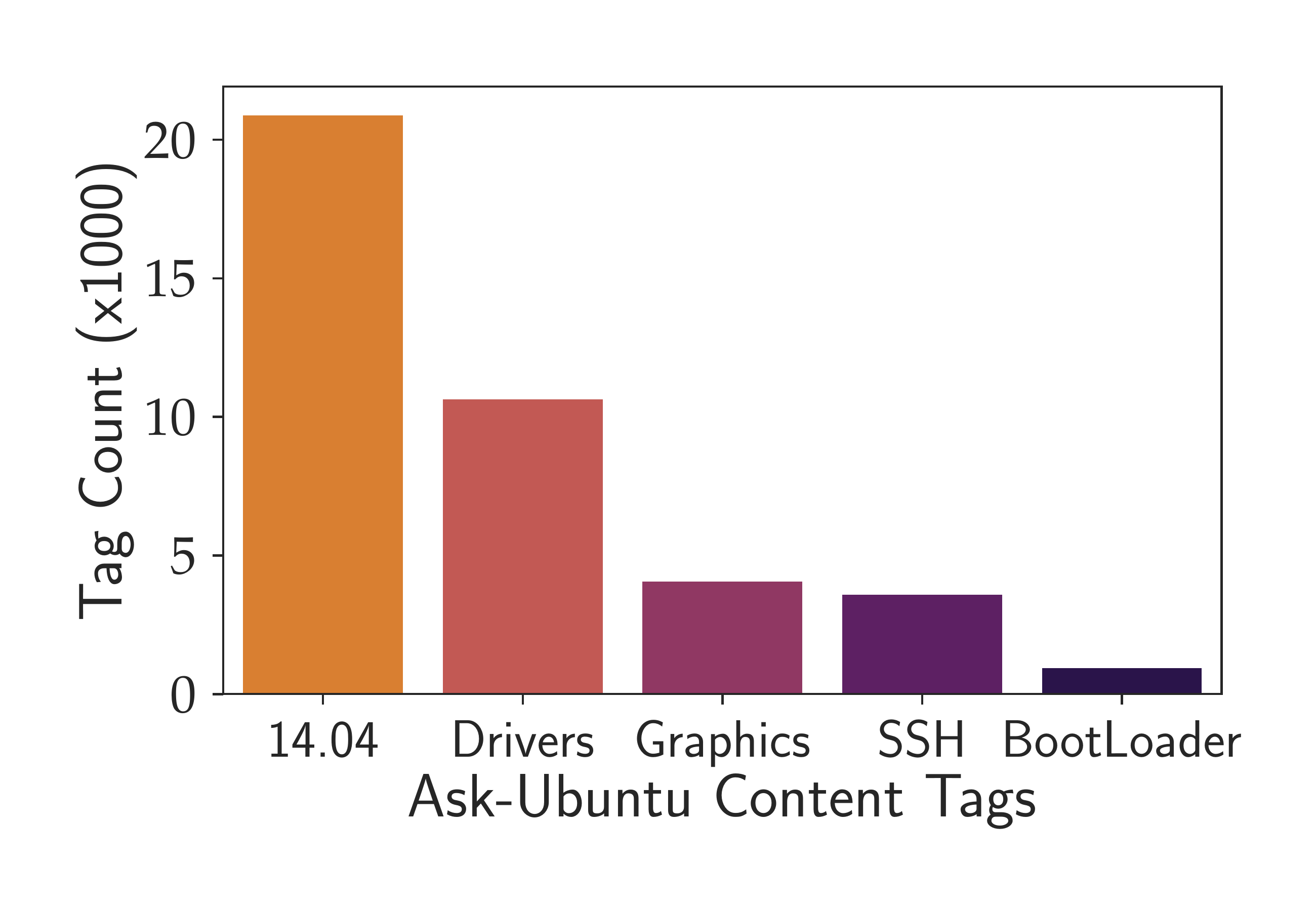}
	\label{skewb}
  }
  \vspace{-14pt}
  \caption{Illustration of (a) Dominant Action Skew and (b) Content Skew in our largest Stack-Exchange, Ask-Ubuntu}
  \vspace{-14pt}
  \label{skew}
 \end{figure}

Behavior skew poses a significant challenge in identifying informative patterns of user engagement with interactive social-media \cite{dir2, qa}. In the Ask-Ubuntu\footnote{https://askubuntu.com/} Q\&A network, most users primarily engage by commenting on posts as seen in \Cref{skewa}. Subject experts who invest most of their time editing or answering questions are relatively infrequent. The extent of behavior skew is compounded by the presence of popular and niche subject areas (\Cref{skewb}). For instance, users who comment on popular topics vastly outnumber those who edit or answer questions on niche subject areas.

\begin{figure}[b]
\vspace{-16pt}
  \subfigure[b][]{
   \centering
   \includegraphics[width=0.466\linewidth, trim={0.2cm 0.1cm 0.36cm 0.5cm},clip]{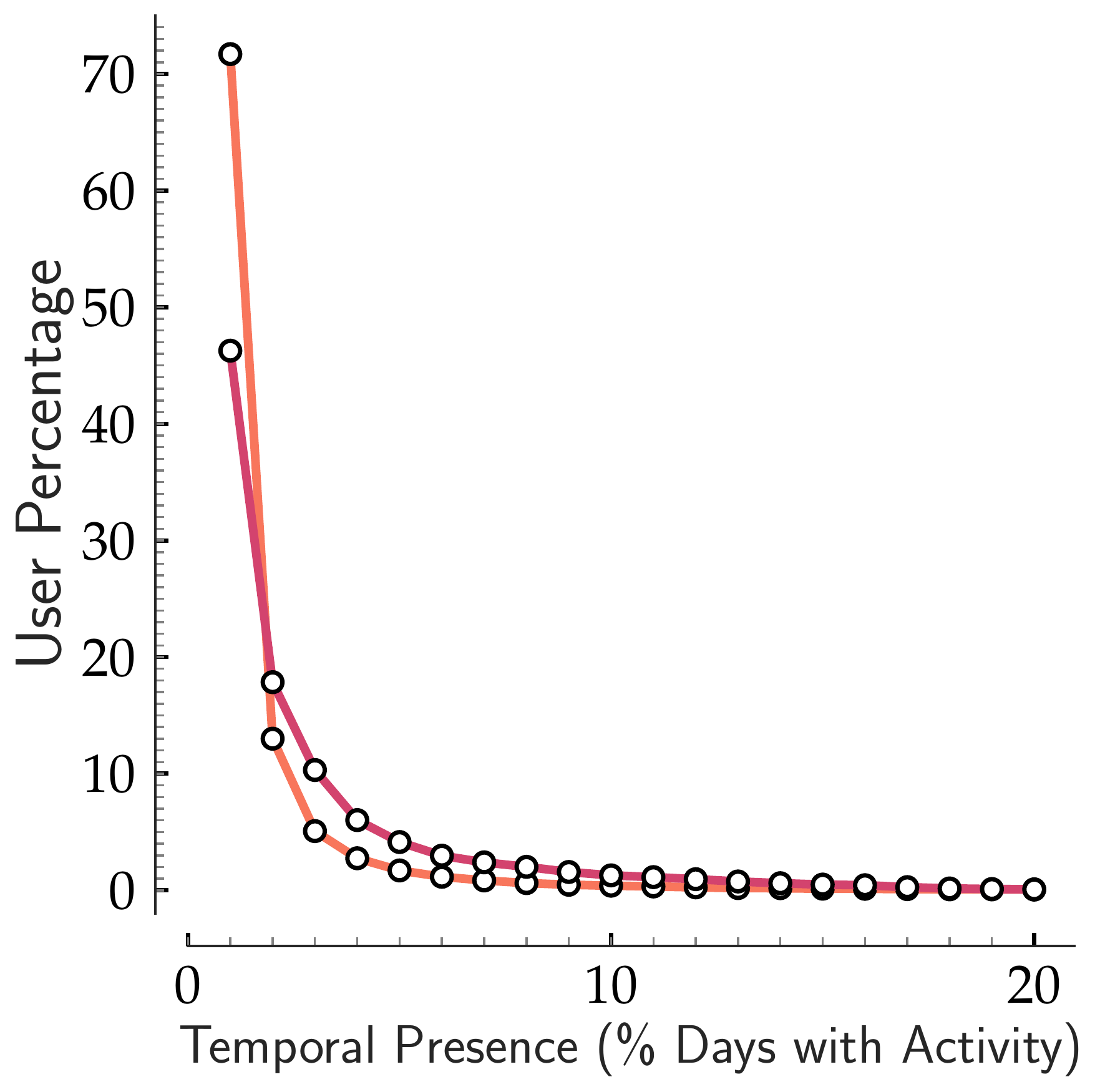}
   \label{sparsitya}
  }
  \subfigure[b][]{
   \centering
   \includegraphics[width=0.466\linewidth, trim={0.2cm 0.1cm 0.36cm 0.5cm},clip]{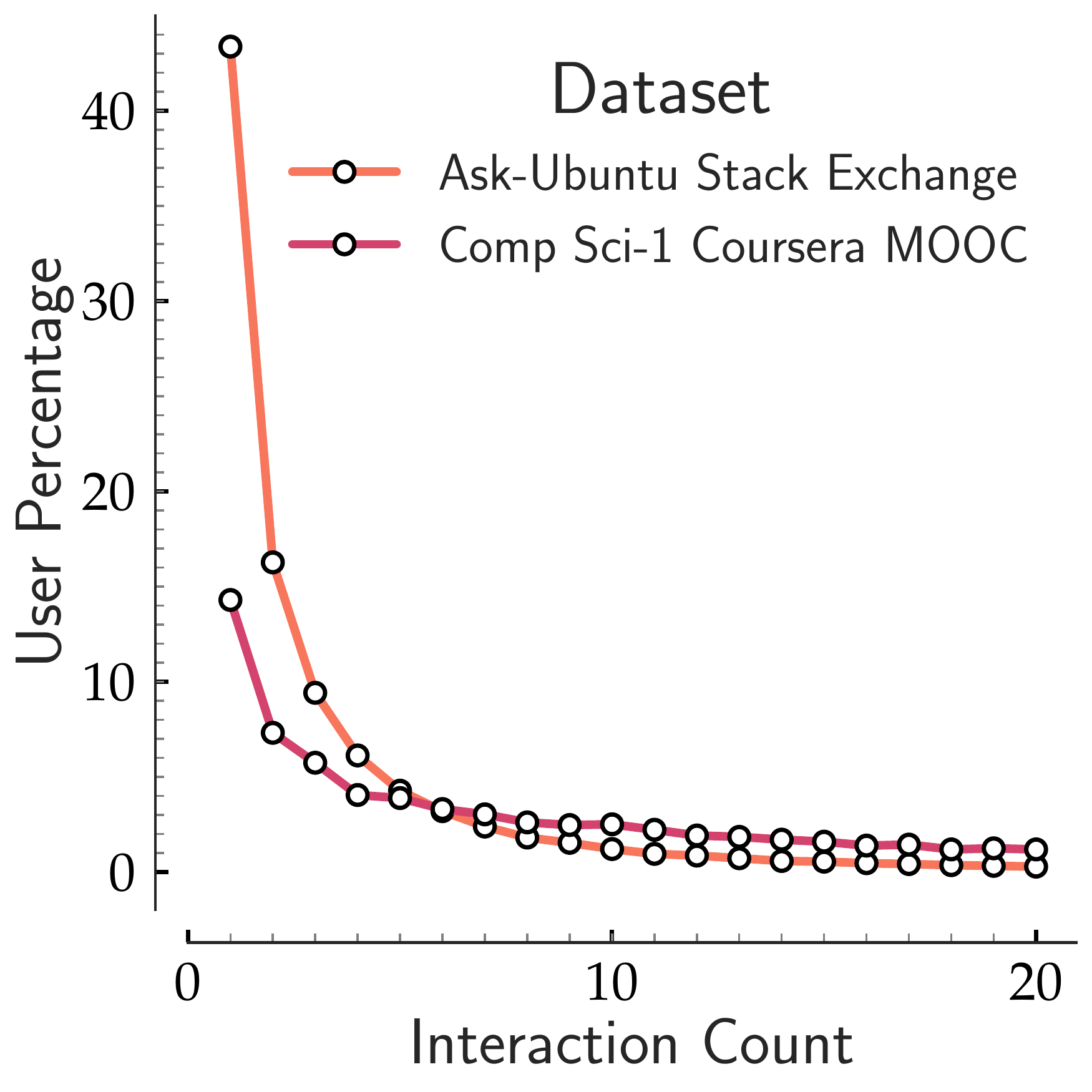}
   \label{sparsityb}
  }
  \vspace{-16pt}
  \caption{Temporal Consistency and User Interaction Volume ($\eta$ $\approx$ 3) are highly skewed in Stack-Exchange/Coursera}
  \vspace{-6pt}
  \label{sparsity}
 \end{figure}

Inconsistent participation and user-level data sparsity are other prominent challenges in most social media platforms \cite{Barabasi1999, fema, sparse1}. In Community Q\&A websites and MOOCs, a minority of participants dominate activity in a classic power-law distribution~\cite{Barabasi1999} as observed in \Cref{sparsitya}. Additionally, an overwhelming majority of users record activity on less than 10\% days of observation in our datasets (\Cref{sparsityb}). Temporal inconsistency in user participation renders evolutionary user modeling approaches \cite{blda, ladfg} ineffective for sparse or bursty participants in social learning platforms.

 Despite several relevant lines of work, including user evolution modeling \cite{ladfg}, behavior factorization \cite{www15}, sparsity-aware tensor factorization \cite{fema} and contextual text mining \cite{blda}, there are few systematic studies addressing these pervasive challenges in modeling user behavior across diverse platforms and applications. Evolutionary activity sequence based user modeling approaches \cite{ladfg} do not explicitly account for sparse or bursty users, and are best suited to temporally consistent user activity.

Matrix factorization methods have been adapted to extract dynamic user representations and account for evolution of user interests. Jiang et al \cite{fema} develop a multi-facet tensor factorization approach for evolutionary analysis, employing facet-wise regularizer matrices to tackle sparsity. \cite{www15} discovers topical interest profiles via simultaneous factorization of action-specific matrices. Quadratic scaling imposes restrictive computational limits on these methods. Generative behavior models are controlled by introducing Dirichlet priors in user profile assignments \cite{blda, spacetime, trirole}. However, this setting is limited in it's ability to model skew, and could merge infrequent behavior profiles with common ones. Furthermore, behaviors learned could be contaminated by the presence of several inactive users.

In contrast to these approaches, we propose to simultaneously partition and profile users in a unified latent framework to adapt to varying degrees of skew and data sparsity in our network. Our user-partioning scheme builds upon preferential attachment models \cite{py, crp} to explicitly discount common activity profiles and favor exploration of diverse user partitions, learning fine-grained representations of behavioral signals. Mutual-enhancement of behavior profiling and user-partitioning can also bridge temporal inconsistencies or sparsity in participant activity. Users exhibiting similar engagement patterns are jointly grouped and profiled within partitions. Furthermore, our latent behavior profiles can be flexibly defined to integrate several facets of user behavior and evolution, hence generalizing to diverse social platforms and applications.

\vspace{2pt}
The main contributions of this paper are:
\begin{itemize}
 \item \textbf{Partitioning and Profiling :} We simultaneously generate flexible user partitions and profile user behavior within partitions in a unified mutually-enhancing latent framework. Our partitioning scheme can adapt to varying levels of behavior skew, effectively uncover fine-grained or infrequent engagement patterns, and address user-level sparsity.
 \item \textbf{Generalizability :} Our model is generalizable to diverse platforms and applications. User profiles can be flexibly defined to integrate several facets of user behavior, social activity and temporal evolution of interests, providing comprehensive user representations.\vspace{2pt}
 \item \textbf{User Evolution :} We formalize our evolutionary profiles to integrate the time-evolving content-action associations observed in user activity and social dynamics.\vspace{2pt}
  \item \textbf{Efficiency :} We provide several optimizations for efficient model inference (see \Cref{sec:Model Inference}) and scale linearly in the size of our datasets compared to quadratic-time scaling in tensor factorization approaches.
\end{itemize}

Extensive experiments over large Coursera\footnote{https://coursera.org} datasets as well as Stack-Exchange websites indicate that our approach strongly outperforms state-of-the-art baselines. We perform three prediction tasks: certificate completion prediction (MOOCs), reputation prediction (Stack Exchange), and behavior distribution prediction. For certificate prediction, we outperform baselines on the AUC measure by 6.26\%-15.97\%; reputation prediction by 6.65\%-21.43\%; behavior prediction MOOCs (12\%-25\%) and Stack-Exchanges (9.5\%-22\%).

On experiments related to activity sparsity, we see magnified gains on participants who offer limited data (10.2\%-27.1\%). We also examine the effects of reducing behavior skew: our approach still outperforms baselines on data with reduced skew. Our scalability analysis shows that the model scales well, and is amenable to a parallel implementation. Finally, we study the effects of model parameters and obtain stable performance over a broad range of parameter values, indicating that in practice our model requires minimal tuning.

We organize the rest of the paper as follows. In the next section, we formalize our data abstraction and user representation. In~\Cref{sec:Platform Description}, we describe the datasets used and in~\Cref{sec:Our proposed Model}, we present our approach.~\Cref{sec:Model Inference} describes a collapsed Gibbs sampler to infer model parameters, and experimental results are presented in~\Cref{sec:Experimental Results}. We discuss related work in~\Cref{sec:related work} and conclude the paper in~\Cref{sec:conclusion}.

\section{Problem Definition}
\label{sec:Problem Definition}
We study networks where participants seek to primarily gain and exchange knowledge (e.g. MOOCs, Stack-Exchanges). In these networks, participants act (e.g. ``post'', ``play video'', ``answer'') on content and communicate with other participants. Content may either be participant-generated (e.g. in a forum), or external (e.g. MOOC lecture). Interactions with content encode latent knowledge and intent of participants - for instance, answering or editing published content is indicative of subject expertise. Furthermore, social exchanges between participants play an important role in profiling their activity.

Let $\mathcal{U}$ denote the set of all participants on the network. These participants employ a finite set of distinct actions $\mathcal{A}$ to interact with content generated from vocabulary $\mathcal{V}$. Atomic participant activity is refered to as an interaction. We define each interaction $d$ as a tuple $d = (a, W, t)$, where the participant performs action $a \in \mathcal{A}$ on content $W=\{w \mid w \in \mathcal{V}\}$ at a normalized time-stamp $t \in [0,1]$. We denote the set of all interactions of participant $u$ as $\mathcal{D}_u$. Thus the entire collection of interactions in the network is given by, $\mathcal{D} = \cup_{u \in \mathcal{U}} \mathcal{D}_u$.

Inter-participant links are represented by a directed multigraph $G=(\mathcal{U}, E)$. A directed labeled edge $(u,v,l) \in E$  is added for each interaction of user $u$, $d_u \in \mathcal{D}_u$(e.g. ``answer'') that is in response to an interaction of user $v$, $d_v \in \mathcal{D}_v$ (e.g. ``ask question'') with edge label $l \in \mathcal{L}$ indicating the specific nature of the social exchange (e.g. ``answer'' \textrightarrow ``question'').

Our model infers a set of latent activity profiles $R$, where each profile $r \in R$ encodes a specific evolutionary pattern of user behavior and social engagement. Observable facets of user behavior, namely $(D_u, L_u), u \in \mathcal{U}$ are drawn from profile $r \in R$ with likelihood $p(D_u, L_u \mid r)$ which we abbreviate as $p(u \mid r)$. Each user is then represented by the vector of likelihoods over latent profiles, i.e. $\mathcal{P}_u = [p(u \mid r) \hspace{2pt}\forall\hspace{2pt} r \in R]$ .

\section{Dataset Description}
\label{sec:Platform Description}
We study datasets from two diverse real-world learning platforms. They encompass rich temporal behavior data in conjunction with textual content, and a community element whereby participants form social ties with each other.
\begin{center}
\begin{table}[t]
\begin{tabular}{llp{0.5\linewidth}}
\toprule
\textbf{Platform} & \textbf{Action} & \textbf{Description} \\
\midrule
\multirow{7}{*}{MOOC} & Play & First lecture segment view \\
& Rewatch & Repeat lecture segment view \\
& Clear Concept & Back and forth movement, pauses \\
& Skip & Unwatched lecture segment \\
& Create Thread & Create a forum thread with a question\\
& Post & Reply to existing threads\\
& Comment & Comment on existing posts\\
\midrule
\multirow{5}{*}{Stack Ex.} & Question & Posting a question\\
& Answer & Authoring answer to a question\\
& Comment & Comment on a question/answer\\
& Edit & Modify posted content\\
& Favorite & Liking posted content\\
\bottomrule
\end{tabular}
\caption{User Action Description (Coursera/Stack-Exchange)}
\label{t1}
\vspace{-42pt}
\end{table}
\end{center}
\textbf{\\Stack-Exchange Q\&A Networks: }Stack-Exchanges are Q\&A websites covering broad domains of public interest. Users interact by asking/answering questions, and editing, liking and commenting on published content (\Cref{t1}). Furthermore, users communicate by reacting to other users' activity, specifically liking and editing content, favorite, and answering questions hence setting up Editing, Liking and Answering links between the pair of users, indicative of their shared interests and knowledge. We apply our model on several Stack-Exchange websites from varied domains (\Cref{d1}\vspace{-8pt}).

\begin{center}
\vspace{-23pt}
\begin{table}[b]
\begin{tabular}{@{}p{0.13\linewidth}p{0.19\linewidth}cccc@{}}
\toprule
\textbf{Platform} & \textbf{Dataset} & \textbf{\#Users} & \textbf{\#Interactions} &\textbf{$\eta_t$} & \textbf{$S_N$}\\
\midrule
\multirow{4}{*}{Coursera}  & Math & 10,796 &162,810 &  -2.90 & 0.69\\
& Nature & 6,940 & 197,367 &-2.43 &0.70 \\
& Comp Sci-1 & 26,542 & 834,439 & -2.51 & 0.67\\
&  Comp Sci-2 & 10,796 & 165,830 & -2.14 & 0.73\\
\midrule
\multirow{10}{*}{Stack-Ex}  & Ask-Ubuntu & 220,365 & 2,075,611 & -2.81 & 0.65\\
& Android & 28,749 & 182,284 & -2.32 & 0.56  \\
& Travel & 20,961 & 277,823 & -2.01 & 0.66\\
& Movies & 14,965 & 150,195 & -2.17 & 0.67\\
& Chemistry & 13,052 & 175,519 & -2.05 & 0.63\\
& Biology & 10,031 & 138,850 & -2.03 & 0.71\\
& Workplace & 19,820 & 275,162 & -2.05 & 0.59\\
& Christianity & 6,417 & 130,822 & -1.71 & 0.64\\
& Comp. Sci. & 16,954 & 183,260 & -2.26 & 0.62\\
& Money & 16,688 & 179,581 &-1.72 & 0.63\\
\bottomrule
\end{tabular}
\caption{Preliminary Analysis of Behavior Skew and Temporal Inconsistency of participant activity in our datasets\vspace{-14pt}}
\label{d1}
\end{table}
\end{center}

\textbf{\\\\Coursera MOOC Platform: } Coursera MOOCs feature a structured learning environment driven by both, lecture content and communication between students and instructors via multiple course forums. Patterns of lecture viewing obtained from video clickstreams provide valuable cues on student learning behavior \cite{mooc1, ladfg}, in addition to forum activity \cite{forum1}. We combine these two sources to define the action set of students (\Cref{t1}). Lecture segment content is extracted from subtitle files. Students engage in social exchanges by commenting on or upvoting content from their peers. We study several MOOC datasets described in \Cref{d1}.\vspace{3pt}

To quantify data sparsity in our datasets, we compute the power-law index ($\eta_t$) that best describes the fraction of users against number of weeks with activity. A more negative index indicates that fewer users are consistently active over time. Behavioral skew can be quantified by grouping participants by their dominant action type (e.g. users who mostly comment), and computing normalized entropy ($S_N$) of the resulting distribution of users across actions. In large Stack-Exchanges such as Ask-Ubuntu, 'Answer' is the dominant action for less than 5\% users while 'Comment' accounts for over 60\% (\Cref{skew}). In MOOCs, 'Play' is the most common action and forum interactions are rare ($\sim$10-15\% participation), resulting in fewer social links. It is interesting to observe that large Stack-Exchanges have more inactive users in comparison to niche domains of discussion (Ask-Ubuntu vs Money, \Cref{d1}).



\section{Our Approach}
\label{sec:Our proposed Model}

\begin{table*}[htbp]
 \centering
 \begin{tabular}{@{}rp{0.75\linewidth}@{}}
  \toprule
  \textbf{Symbol}                                                                  & \textbf{Description}                                                                                                                           \\
  \midrule
  $\mathcal{D, U, A, V}$                                                           & Set of all content interactions, users, actions and content vocabulary                                                                             \\
  $\mathcal{D}_{u}, \mathcal{L}_u$                                                                & Content interactions and social links observed for user $u \in \mathcal{U}$                                                                                             \\
  $d = (a,W,t)$                                                 & Interaction involving action $a$ on content $W$ at time $t$                             \\
   $(s,u,l), (u,y,l') \in \mathcal{L}_u; l,l' \in \mathcal{L}$                                                                  & l-label inward link from source s, l'-label outward link to target y; $\mathcal{L}$ - predefined link label set                                                         \\
  $R$, $K$                                                                             & The set of evolution profiles, and the set of behavior topics                                                                                  \\
  $\phi_{k}^{\mathcal{V}}, \phi_{k}^{\mathcal{A}}$                                 & Multinomial word and action distributions in behavior $k \in K$                                                                                  \\
  $\phi_{r}^{K}$                                                                   & Multinomial distribution over behaviors for profile $r \in R$                                                                                  \\
  $\alpha_{rk}, \beta_{rk}$                                                        & Parameters of the temporal beta distribution for behavior $k$ in profile $r$                                                                      \\
  $\phi_{r, r'}^{\mathcal{L}}$                                                                & Multinomial distribution over link labels for links directed from profile $r$ to $r'$                                     \\
  $\gamma, \delta, G_{0}$                                                               & Scale parameter, discount parameter and base distribution of Pitman-Yor process                                                                \\
  $a, n_{a}, r_{a}; N, A_{r}$                                                         & Table ID, \# seated users, profile served on it; \# total seated users and \# tables serving profile $r \in R$                                         \\
  $\alpha_{\mathcal{V}}, \alpha_{\mathcal{A}}, \alpha_{K}, \alpha_{\mathcal{L}}$ & Dirichlet-Multinomial priors for $\phi_{k}^{\mathcal{V}}$, $\phi_{k}^{\mathcal{A}}$, $\phi_{r}^{K}$ and $\phi_{r, r'}^{\mathcal{L}} $  $\forall$ $k \in K$ \& $r,r' \in R$ \\
  \bottomrule
 \end{tabular}
 \caption{Notations used in this paper}
 \label{tnotation}
 \vspace{-24pt}
\end{table*}
In this section, we motivate our behavior profiling framework (\Cref{sub:Motivation}) proceeding in two simultaneous mutually-enchancing steps. Discovering diverse homogenous partitions of users in the network (\Cref{sub:Skew Aware User Partitioning}), and learning latent evolutionary profiles characterizing facets of their content interactions and social exchanges (\Cref{sub:Model Constituents}). \vspace{-5pt}
\subsection{Motivation}
\label{sub:Motivation}
We develop our profiling framework with two key objectives. First, to account for behavior skew (participants are unevenly distributed across varying behavior patterns) as well as temporal inconsistency and sparsity in user-level data. Second, to learn informative evolutionary profiles simultaneously characterizing engagement with content and social exchanges between users.

Modeling inherent behavior skew necessitates the development of profiling approaches that adapt to the observed data and learn informative representations of user activity. Generative behavior models are traditionally controlled by introducing Dirichlet priors in the topic assignment process \cite{blda, spacetime, trirole}. However, this setting is limited in it's ability to model inherent topical skew, where some outcomes significantly outnumber others. In the context of user behavior, it is necessary to explicitly account for the presence of skew in the proportions of behavior patterns and effectively separate users to learn discriminative evolutionary profiles of activity. There are two key advantages to mutually enhancing user paritioning and profile learning over conventional profile assignments:
\begin{itemize}
\item \textbf{Tackling Behavior Skew} : Conventional topic assignments tend to merge infrequent behavior patterns with common ones, resulting in uninformative profiles. Our approach explicitly discounts common profiles and favors diverse user partitions. Profile variables assigned to these partitions learn informative representations of infrequent behaviors.\vspace{3pt}
\item \textbf{Temporal Inconsistency and Sparsity} : Our profile assignment process enforces common profiles across inconsistent and active users within homogenous partitions. As our inference process converges, users with limited data are probabilistically grouped with the most similar users based on available interaction data.
\end{itemize}
We now proceed to formalize our user partitioning scheme.

\subsection{Skew Aware User Partitioning}
\label{sub:Skew Aware User Partitioning}
We first introduce a basic preferential attachment model based on the Pitman-Yor process \cite{py} to generate skewed partitions of integeres. We proceed to develop our profile-driven user partitioning scheme, building upon the Chinese Restaurant perspective \cite{crp} of the Pitman-Yor process to group similar users within partitions. Our approach explicitly discounts common behavior profiles to generate diverse user partitions and learn subtle variations and infrequent behavior patterns in the network. Additionally, it jointly profiles sparse and temporally inconsistent users with best-fit partitions.
\subsubsection{Basic Preferential Attachment}
The Pitman-Yor process \cite{py} (generalization of the Dirichlet process \cite{dp}) induces a distribution over integer partitions, characterized by a concentration parameter $\gamma$, discount parameter $\delta$, and a base distribution $G_0$. An interpretable perspective is provided by the Chinese Restaurant seating process \cite{crp} (CRP), where users entering a restaurant are partitoned across tables. Each user is either seated on one of several existing tables $[1, \ldots, A]$, or assigned a new table $A+1$ as follows,

\begin{equation}
 \label{eq:basicCRP}
 p(a | u) \propto
 \begin{cases}
  \frac{n_{a}-\delta}{N + \gamma}, \quad a \in [1, A], \text{ existing table} \\[5pt]
  \frac{\gamma+ A\delta}{N + \gamma}, \quad a=A+1, \text{ new table}          \\
 \end{cases}
\end{equation}
where $n_{a}$ is the number of users seated on existing tables $a \in [1,A]$, $A+1$ denotes a new table, and $N = \sum_{a \in [1,A]}n_{a}$ is the total number of participants seated across all tables.~\Cref{eq:basicCRP} thus induces a preferential seating arrangement proportional to the current size of each partition. The concentration ($\gamma$) and the discount ($\delta$) parameters govern the formation of new tables.

This simplistic assignment thus captures the ``rich get richer'' power-law characteristic of online networks~\cite{Barabasi1999}. However a significant drawback is it's inability to account for similarities of users seated together. Our approach enforces a profile-aware seating arrangment to generate homogenous user partitions.

\subsubsection{Profile-Driven Preferential Attachment}
Let us now assume the presence of a set of evolutionary profiles, $R$ describing temporal patterns of user engagement and their social ties. Each user u $\in \mathcal{U}$ is associated with a set of time-stamped interactions $D_u$, and social links $L_u$. The likelihood of generating these observed facets via profile $r \in R$ is given by p($D_u, L_u \mid r$), which we abbreviate to p($u \mid r$).

Thus, to continue the restaurant analogy above, we ``serve'' a table-specific profile $r_{a} \in R$ to  participants seated on each table $a \in [1,A]$. When we seat participant $u$ on a new table $A+1$, a profile variable $r_{A+1} \in R$ is drawn on the new table to describe $u$,

\begin{equation*}
 r_{A+1} \sim p(u \mid r)  p(r)
\end{equation*}

where $p(r)$ is parameterized by the base distribution $G_0$ of the Pitman-Yor process, acting as a prior over the set of profiles. We set $p(r)$ to a uniform distribution to avoid bias. A user $u$ in our model is thus seated on table $a$ as follows,
\begin{equation}
 \label{eq:updatedCRP}
 p(a | u) \propto
 \begin{cases}
  \frac{n_{a}-\delta}{N + \gamma} \times p(u \mid r_{a})   ,                           & a \in [1, A], \\[5pt]
  \frac{\gamma+A \delta}{N + \gamma} \times \frac{1}{|R|} \sum_{r \in R} p(u \mid r) , & a=A+1.        \\
 \end{cases}
\end{equation}
Thus, the likelihood of assigning a specific profile $r$ to participant $u$, $p(r|u)$ is obtained by summing up over the likelihoods of being seated on existing tables serving $r$, and the likelihood of being seated on a new table $A+1$ with the profile draw $r_{A+1}=r$,
\begin{align}
 \label{eq:profiledist}
 p(r \mid u) = & (\sum_{a \in [1,A], r_{a} = r} \frac{n_{a} - \delta}{N+\gamma}p(u \mid r)) +                                         \frac{1}{|R|}  \cdot \frac{\gamma+A\delta}{N + \gamma} p(u \mid r) \\
 =            & \left ( \frac{N_{r} - A_{r} \delta}{N + \gamma}  + \frac{\gamma+A\delta}{|R|(N + \gamma)}   \right ) p(u \mid r)
\end{align}
where $A_{r}$ is the number of existing tables serving profile $r$ and $ N_{r} $ is the total number of participants seated on tables serving $r$. \vspace{2pt}

\textbf{Discount Factor : }The extent of skew is jointly controlled by both, the number of users exhibiting similar activity patterns, encoded by $p(u \mid r)$ as well as the setting of the discount parameter $\delta$. Common profiles are likely to be drawn on several tables, thus their probability masses are discounted by the product $A_{r} \delta$ in \Cref{eq:profiledist}. A higher setting of $\delta$ favors exploration by seating users on new tables and generating diverse partitions, learning subtle variations in profiles rather than merging them. \vspace{3pt}

\textbf{Temporal Inconsistency : }Users offering limited evidence are likely to be assigned to popular profiles that well explain their limited interaction data. Our user partitioning scheme enforces a common profile assignment across users sharing a partition, thus ensuring proximity of their likelihood distributions over the latent space of the inferred evolutionary profiles.
\newpage

\begin{algorithm}[t]
 \caption{Profile-Driven Preferential Attachment}
 \begin{algorithmic}[1]
  \For{each user $u \in \mathcal{U}$}
  \State Sit at existing table $a \in [1,A]$ $\propto \frac{n_{a}-\delta}{N + \gamma} \times p(u \mid r_a)$
  \State Sit at new table $A+1$ $\propto \frac{\gamma+A\delta}{N + \gamma} \times (\sum_{r \in R}p(u \mid r) \times \frac{1}{|R|})$
  \If{$A+1$ is chosen}
  \State Draw $r_{A+1}$ $\propto$ $p(u \mid r) \times \frac{1}{|R|},  r_{A+1} \in R$
  \EndIf
  \EndFor
 \end{algorithmic}
 \label{alg:crpmap}
\end{algorithm}
\setlength{\textfloatsep}{4pt}
Simplistic prefential assignment (\Cref{eq:basicCRP}) can be interpreted as a specialization of our model where all evolutionary profiles $r \in R$ are equally likely for every user. Our model effectively generalizes \Cref{eq:basicCRP} to generate partitions of typed entities rather than integers. The resulting seating arrangement in our model can be shown to be exchangeable similar to that in \cite{crp} and hence amenable to efficient inference. Our partitioning approach can be extended to several diverse applications, depending on the precise formulation of  p($u \mid r$). In the next subsection, we formalize our definition of evolutionary profiles $r \in R$.

\subsection{Evolutionary Activity Profiling}
\label{sub:Model Constituents}
We now formalize the notion of latent behaviors and evolutionary activity profiles.
A behavior (or a behavioral topic) $k \in K$ is jointly described by $\phi_{k}^{\mathcal{V}}$, a $|\mathcal{V}|$ dimensional multinomial distribution over words, and  $\phi_{k}^{\mathcal{A}}$, a  $|\mathcal{A}|$ dimensional multinomial distribution over the set of actions.
The combined probability of observing an action $a$ (e.g. ``play'', ``post'') on content $W = \{w \mid w \in \mathcal{V}\}$ (e.g. a sentence on ``Probability'') conditioned on behavior $k$ can be given by,
\begin{equation}
 p(a, W \mid  k) \propto \phi_{k}^{\mathcal{A}}(a) \prod_{w \in W}^{} \phi_{k}^{\mathcal{V}}(w).
 \label{pdk}
\end{equation}
Each observed interaction is assumed to be drawn from a single behavior $k \in K$, thus learning consistent action-content associations.

Evolutionary profiles are temporal mixtures over behaviors. We describe each activity profile $r \in R$ jointly by a $K$ dimensional multinomial-distribution parameterized by $\phi_{r}^{K}$ over the $K$ latent behaviors, and a \textit{Beta} distribution specific to each behavior $k \in K$, over normalized time $t \in [0,1]$, parameterized by $\{\alpha_{rk}, \beta_{rk}\}$. Each component of the multinomial distribution $\phi_{r}^{K}(k)$ is the likelihood of observing behavior $k$ in profile $r$. The $\alpha_{rk}, \beta_{rk}$ parameters of the \textit{Beta} distributions capture the temporal trend of behavior $k$ within profile $r$.

We draw an interaction ${d = (a, W, t)}$ within profile $r$ by first drawing a behavior $k$ in proportion to $\phi_{r}^{K}(k)$. We then draw action $a$ and content $W$ conditioned on behavior $k$, and time $t$ conditioned on both $r$ and $k$. Thus, the likelihood of observing interaction $d$ in profile $r$, $p(d \mid r)$ is obtained by marginalizing over behaviors,
\begin{equation}
 p(d \mid r) \propto \sum_{k} \phi_{r}^{K}(k) \times p(a, W \mid  k) \times p(t \mid r,k).
 \label{eqx}
\end{equation}
where $p(a, W \mid  k)$ is computed as in~\Cref{pdk} and $p(t \mid r,k)$ through the corresponding \textit{Beta} distribution $Beta(t; \alpha_{rk}, \beta_{rk})$,
\begin{equation}
 p(t \mid r,k) = \frac{t^{\alpha_{rk} - 1} (1-t)^{\beta_{rk} - 1}}{B(\alpha_{rk},\beta_{rk})}.
 \label{time}
\end{equation}

The \textit{Beta} distribution offers us flexibility in modeling temporal association. Prior behavior models~\cite{fema, ladfg} used static time slicing to describe user evolution. Choosing an appropriate temporal granularity is challenging. A single granularity may be inadequate to model heterogeneous user activity. Since we analyze behavior data recorded over finite intervals, the parameterized Beta distribution is capable of learning flexible continuous variations over normalized time-stamps via parameter estimation \cite{tot}.

We can now compute the likelihood of observing the entire set of interactions $\mathcal{D}_u$ of a user $u \in \mathcal{U}$ conditioned on profile $r$ as,
\begin{equation}
 p(D_{u} \mid r) \propto \prod_{d \in \mathcal{D}_u} p(d \mid r)
 \label{ur}
\end{equation}
The above process is summarized in~\Cref{alg:dup}. In addition to interaction set $\mathcal{D}_u$, we now proceed to exploit inter-participant link structure in the construction of activity profiles.
\begin{center}
\begin{figure}[b]
 \includegraphics[width=0.92\linewidth, trim={7.8cm 0cm 6.8cm 0.2cm},clip]{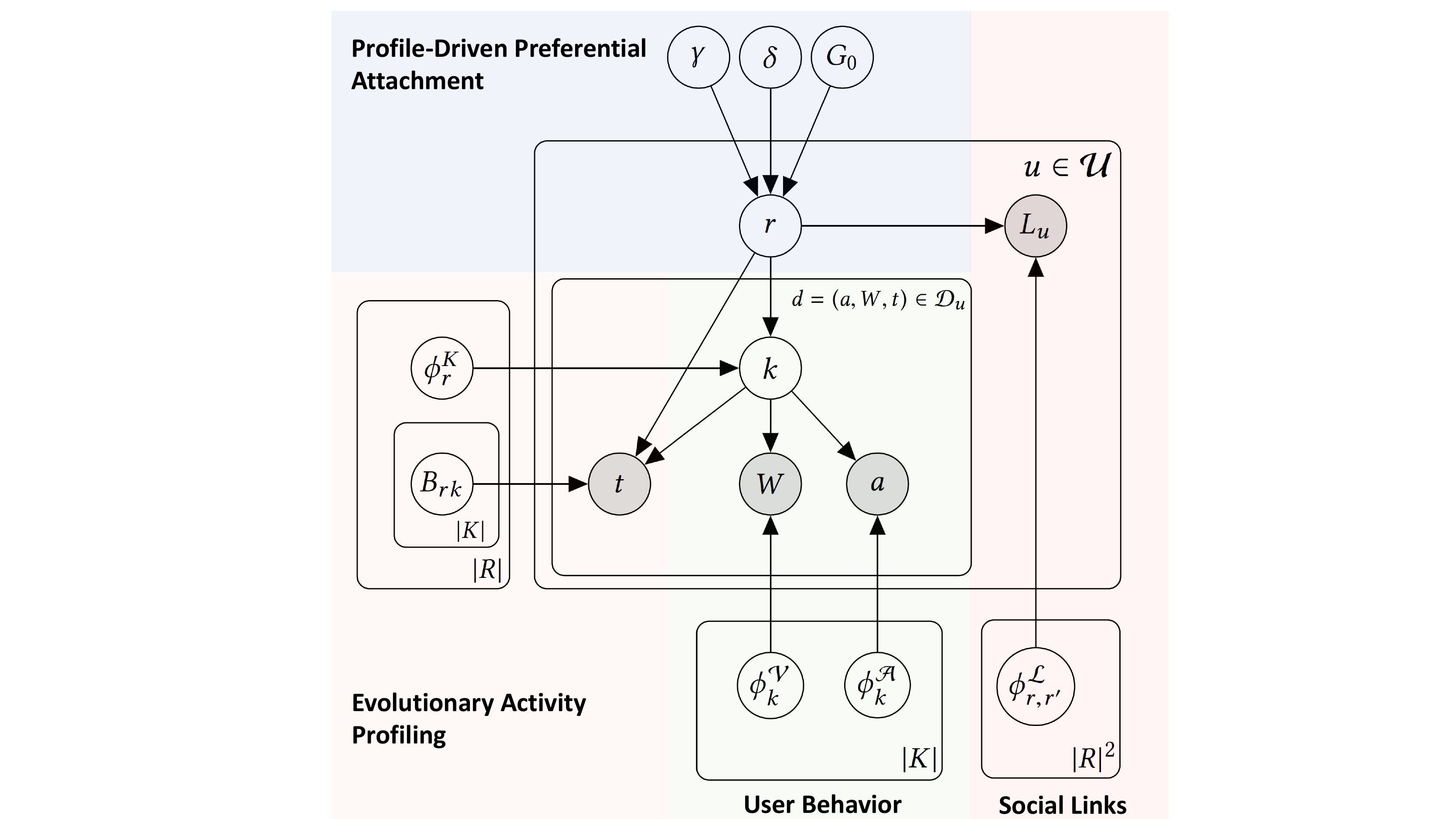}
 \caption{Graphical model illustrating our simultaneous partitioning and profiling framework}
 \label{crpmap}
\end{figure}
\end{center}
\vspace{-15pt}
\subsubsection{Modeling Participant Links}
\label{sub:Modeling Participant Links}
We create a directed multi-graph between pairs of profiles to incorporate the nature of participant ties. Edge labels $l \in \mathcal{L}$ describe the nature of exchanges (e.g. ``question'', ``answer'') connecting users who perform them, and the directionality of the exchange is indicative of the implicit social relationship between their profiles. We thus associate a multinomial distribution parameterized by $\phi_{r,r'}^{\mathcal{L}}$ between each ordered pair of profiles $r, r'$, setting up $|R|^2$ distributions in all.

The network structure also enhances modeling for participants with inconsistent interaction data to take advantage of more extensive interaction data of their neighbors. The probability $p(L_{u} \mid r_{u})$ of the set of links $L_{u}$ associated with user $u$ is proportional to the likelihoods of independently observing each link, based on the current profile assignment $r_u \in R$, and the participant profiles of users linked to u,
\begin{equation}
 p(L_{u} \mid r_{u}) \propto \prod_{(s,u,l) \in L_{u}} \phi_{r_s,r_u}^{\mathcal{L}}(l) \times \prod_{(u,y, l) \in L_{u}}\phi_{r_u,r_y}^{\mathcal{L}}(l),
 \label{eqxl}
\end{equation}
where $\phi_{r_s,r_u}^{\mathcal{L}}(l)$ is the likelihood of an $l$-labeled inward link to user $u$ emerging from a source user $s$ with profile $r_s$, and $\phi_{r_u,r_y}^{\mathcal{L}}(l),$ is that of the analogous outward link to user $y$. We thus encode observed social links as implicit relationships of the respective evolutionary profiles.

We combine social ties $L_{u}$ and content interactions $D_{u}$ (\cref{eqxl,ur}), to compute the joint conditional probability $p(u \mid r)$,
\begin{equation}
 P(u \mid r) \propto p(D_{u} \mid r) \times p(L_{u} \mid r).
 \label{upfinal}
\end{equation}
The above equation provides the likelihood of describing user $u \in \mathcal{U}$ with a chosen profile $r \in R$. The generative process correspending to \cref{upfinal} is summarized by ~\Cref{alg:dup}.
\begin{algorithm}[t]
 \caption{Generative process for drawing facets $\mathcal{D}_u$, $L_{u}$ from profile $r \in R$ assigned to the partition containing user $u \in \mathcal{U}$}
 \begin{algorithmic}[1]
  \For{each behavior $k \in K$}
  \State Draw word distribution $\phi_{k}^{\mathcal{V}}$ $\sim$ $Dir(\alpha_{\mathcal{V}})$
  \State Draw action distribution $\phi_{k}^{\mathcal{A}}$ $\sim$ $Dir(\alpha_{\mathcal{A}})$
  \EndFor
  \For{each profile $r \in R$}
  \State Draw distribution over behaviors, $\phi_{r}^{K}$ $\sim$ $Dir(\alpha_K)$
  \For{each profile $r' \in R$}
  \State Choose link distribution $\phi_{r,r'}^{\mathcal{L}}$ $\sim$ $Dir(\alpha_{\mathcal{L}})$
  \EndFor
  \EndFor
  \For{each behavior interaction $d = (a,W,t) \in \mathcal{D}_u$}
  \State Choose behavior $k$ $\sim$ $Multi(\phi_{r}^{K})$
  \For{word $w$ $\in$ $W_d$}
  \State Draw $w$ $\sim$ $Multi(\phi_{k}^{\mathcal{V}})$
  \EndFor
  \State Draw action $a$ $\sim$ $Multi(\phi_{k}^{\mathcal{A}})$
  \State Draw normalized time $t$ $\sim$ $Beta(\alpha_{rk}, \beta_{rk})$
  \EndFor
  \For{each inward link (s,u,l) $\in L_u$}
  	\State Let $r_s$ denote source user profile
	\State Draw $(s,u,l) \sim Multi(\phi_{r_s,r}^{\mathcal{L}})$
  \EndFor
  \For{each outward link (u,y,l) $\in L_u$}
  	\State Let $r_y$ denote target user profile
	\State Draw $(u,y,l) \sim Multi(\phi_{r,r_y}^{\mathcal{L}})$
  \EndFor
 \end{algorithmic}
 \label{alg:dup}
\end{algorithm}
%
%
%
%
%
%
%
%
%

In this section, we motivated our partitioning and profiling framework to tackle the issues of skew and sparsity in social learning environments. Next, we describe a collapsed Gibbs-sampling approach \cite{cgs} for model inference, where we iteratively sample user seating arrangements and update profile parameters to reflect the resulting set of user partitions, until mutual convergence.

\section{Model Inference}
\label{sec:Model Inference}
In this section we describe a collapsed Gibbs-sampling \cite{cgs} approach for model inference, an analysis of it's computational complexity, and propose an efficient parallel batch-sampler to scale to large datasets.\vspace{-5pt}
\subsection{Inference via Gibbs Sampling}
\label{subsec1}
We exploit the widely used Markov-Chain Monte-Carlo(MCMC) algorithm, collapsed Gibbs-sampling, to sample user seating and learn profiles by iteratively sampling the latent profile variable $r_u$ for each user $u \in \mathcal{U}$, latent behavior-topic assignments $k_d$ for interactions $d \in \mathcal{D}_u$, and table $a_u$ serving sampled profile $r_u$, on which user $u$ is seated.
\begin{center}
\vspace{-5pt}
 \begin{table}[htbp]
  \begin{tabular}{lp{0.7\linewidth}}
   \toprule
   \textbf{Symbol}                         & \textbf{Description}                                                                                                                                    \\
   \midrule
   $n_{k}^{(w)}, n_{k}^{(a)}, n_{k}^{(.)}$ & Number of times word $w$, action $a$ were assigned to topic k, and respective marginals                                                       \\
   $n_r^{(k)}, n_r^{(.)}$                    & Number of interactions of users in $r$ assigned topic k, total interactions of all users in $r$ \\
   $n_{r,r'}^{(l)}, n_{r,r'}^{(.)}$                            & Number of $l$-label links across users in profile $r$ with $r'$, and total links between $r$ and $r'$                                                             \\
   \bottomrule
  \end{tabular}
  \caption{Gibbs-sampler count variables}
  \label{gibbsnotation}
  \vspace{-26pt}
 \end{table}
\end{center}
\subsubsection{Initialization:}
Randomized initializtion of user partitions and corresponding profile assignments could result in longer convergence times. We speed-up convergence by exploiting content tags and action distributions of users to generate a coherent initial seating arrangement. Users with similar action and content tag distributions are seated together to form homogenous partitions.
\subsubsection{Sampling User Partitions:}
The likelihood of generating interaction $d = (a,W,t) \in \mathcal{D}_{u}$ from behavior $k \in K$ (\Cref{pdk}) can be given by,
\begin{equation}
 p(a,W \mid k) \propto  \frac{n_{k}^{(a)} + \alpha_{\mathcal{A}}}{n_{k}^{(.)} + |\mathcal{A}|\alpha_{\mathcal{A}}} \times \prod_{w \in W}\frac{n_{k}^{(w)} + \alpha_{\mathcal{V}}}{n_{k}^{(.)} + |\mathcal{V}|\alpha_{\mathcal{V}}}
 \label{dk}
\end{equation}
Thus the likelihood of observing interaction $d = (a,W,t)$ in profile $r \in R$ (\Cref{eqx}) is,
\begin{equation}
p(d \mid r) \propto \sum_{k \in K} \frac{n_{r}^{k} + \alpha_K}{n_{r}^{(.)} + |K|\alpha_K} \times p(a,W \mid k) \times p(t \mid r,k)
 \label{dr}
\end{equation}
where $p(t \mid r,k)$ is computed as in \cref{time}. Link likelihood for source profile $p$, target $p'$ and label $l$ is computed as,
\begin{equation}
\phi_{p,p'}^{\mathcal{L}}(l) = \frac{n_{p, p'}^ l + \alpha_{\mathcal{L}}}{n_{p, p'}^{(.)} + |\mathcal{L}|\alpha_{\mathcal{L}}}
\label{lr}
\end{equation}
Thus $p(u \mid r), u \in \mathcal{U}$ (\cref{ur}) can be obtained as the product of \cref{dr} over $d \in \mathcal{D}_u$ and \cref{lr} over $L_u$ respectively. Given $p(u \mid r)$ we can sample profile $r_u$ for user $u$ as in \cref{eq:profiledist},
\begin{equation}
 P( r_u = r \mid u, a_{-u}, r_{-u}, k_{-u}) \sim \left ( \frac{N_{r} - A_{r} \delta}{N+\gamma}  + \frac{\gamma+A\delta}{|R|(N+\gamma)}   \right ) p(u \mid r)
 \label{ul}
\end{equation}
where $a_{-u},r_{-u},k_{-u}$ indicate the seating and profile assignments of all other users, and the behavior assignments for their interactions. Behavior assignments for each interaction $d \in \mathcal{D}_u$ are sampled in proportion to \cref{dr} with $r = r_u$, the chosen profile for $u$, and the user is seated either on an existing table $a \in [1, A]$ serving $r_u$, or new table $A+1$ with $r_{A+1} = r_u$,
$$
\begin{cases}
 a \in [1,A] \propto \frac{n_{a}-\delta}{N+\gamma}    \text{ if } r_a = r_u, \text{else 0}                  \\
 \text{New table } A+1 \propto \frac{\gamma+(\delta \times A)}{N+\gamma} \times \frac{1}{|R|}, r_{A+1} = r_u \\
\end{cases}
$$
Note that $N = |\mathcal{U}|-1$, i.e. all users except $u$.

\subsubsection{Parameter Estimation: }All counts (\Cref{gibbsnotation}) corresponding to previous behavior and profile assignments of $u$ are decremented and updated based on the new assignments drawn. At the end of each sampling iteration, Multinomial-Dirichlet priors $\alpha_{\mathcal{V}}$, $\alpha_{\mathcal{A}}$, $\alpha_{K}$ and $\alpha_{\mathcal{L}}$ are updated by Fixed point iteration \cite{fpi} and profile parameters $(\alpha_{rk}, \beta_{rk})$ are updated by the method of moments \cite{tot}.

All time-stamps are rounded to double-digit precision and values of $p(t \mid r,k)$ $\forall$ $t \in [0,1], r \in R, k \in K$ are cached at the end of each sampling iteration. This prevents $R \times K$ scaling for $p(u \mid r)$ in \Cref{ul} by replacing computations with fetch operations. Pitman-Yor parameters can be estimated via auxiliary variable sampling with hyperparameters set to recommended values in \cite{pytm, teh}.
\subsection{Computational Complexity}
In each Gibbs iteration, \Cref{dk,dr} involve $|\mathcal{D}| \times (K+R)$ computations. \Cref{ul} requires an additional $|\mathcal{U}| \times R$ computations. $R \times K$ scaling for $p(u\mid r)$ in \Cref{ul} is prevented by restricting temporal precision as described in \cref{subsec1}. The first product term of \Cref{ul} is cached for each $r \in R$, and updated only when tables of profile $r$ are altered.
\\
On the whole, our algorithm is linear in $|\mathcal{D}| + |\mathcal{U}|$, scaled by $R + K$ in both time and space complexity (results in \Cref{fig:memtime}). We efficiently scale to massive datasets by parallelizing our algorithm across users via batch-sampling, described in the next subsection.

\subsection {Parallelizing Model Inference}
The Gibbs sampler described above samples each user's seating $P( r_u = r \mid u, a_{-u}, r_{-u}, k_{-u})$ conditioned on all other users, which necessitates iteration over $\mathcal{U}$. Instead, seating arrangements could be simultaneously sampled in batches $U \subset \mathcal{U}$ conditioned on all users outside the batch, i.e. $P( R_U = R \mid U, a_{\mathcal{U}-U}, r_{\mathcal{U}-U}, k_{\mathcal{U}-U})$.

Batch sampling is most efficient when each batch $U \subset \mathcal{U}$ is chosen such that users $u \in U$ entail comparable computational loads. We approximate computation load for $u \in \mathcal{U}$ $\propto$ $|\mathcal{D}_u| + |\mathcal{L}_u|$ to decide apriori batch splits for sampling iterations.

All assignment counts can be updated at the end of the sampling process for one batch. Note that social links between users in a given batch $U$ are ignored since their profiles are drawn simultaneously. However, in practice the batch-size is a small value in comparison to $|\mathcal{U}|$, thus rendering this loss to be negligible.


\section{Experimental Results}
\label{sec:Experimental Results}
In this section, we present our experimental results on datasets from MOOCs as well as Stack-Exchange. We first introduce the set of competing baselines. Then in~\Cref{sub:Prediction Tasks}, we discuss prediction tasks used to evaluate the different behavioral representation models. In~\Cref{sub:Effects of data sparsity}, we demonstrate the impact of data sparsity on prediction tasks, and in~\Cref{sub:Effects of behavioral skew}, we discuss the effects of behavior skew. Next, we present results on the parameter sensitivity and scalability of our model in~\Cref{sub:Parameter Sensitivity Analysis,sub:Scalability Analysis} and conclude with a discussion on the limitations of our approach.
%
\subsection{Baseline Methods}
\label{sub:Baseline Methods}
We compare our model (CMAP) with three state-of-the-art behavior models and two standard baselines.
\begin{description}
 \item \textbf{LadFG} \cite{ladfg}: LadFG is a dynamic latent factor model which uses temporal interaction sequences and demographic information of participants to build latent representations. We provide LadFG action-content data from interactions and all available user demographic information.
 \item \textbf{BLDA} \cite{blda}: BLDA is an LDA based extension to capture actions in conjunction with text. It represents users as a mixture over latent content-action topics.
 \item \textbf{FEMA} \cite{fema}: FEMA is a multifaceted sparsity-aware matrix factorization approach which employs regularizer matrices to tackle sparsity. Facets in our datasets are users, words and actions. We set User-User and Word-Word regularizers to link and co-occurrence counts respectively. We could not run FEMA on Ask-Ubuntu and Comp Sci-1 datasets owing to very high memory and compute requirements. Regularizer matrices in FEMA scale as $|\mathcal{U}|^2$, whereas our model scales as $|\mathcal{U}|$.
 \item \textbf{DMM (Only text)} \cite{dmm}: We apply DMM to the textual content of all interactions to learn topics. Users are represented by the probabilistic proportions of learned topics in their interaction content.
 \item \textbf{Logistic Regression Classifier (LRC)} \cite{lrc}: It uses logistic regression to train a classifier model for prediction. Input features include topics(obtained from DMM) that the user interacts with and respective action proportions for each topic.
\end{description}
We initialize Dirichlet priors for $\phi_{k}^{\mathcal{V}}, \phi_{k}^{\mathcal{A}}, \phi_{r}^{K}$ and $\phi_{r,r'}^{\mathcal{L}}$ following the common strategy~\cite{dir1,dir2} ($\alpha_{X} = 50/|X|, X=\{\mathcal{A}, \mathcal{L}, K\}$, and $\alpha_{\mathcal{V}} = 0.01$) and all $\textit{Beta}$ parameters $\alpha_{rk},\beta_{rk}$ are initialized to 1. We found CRP parameter initialization at $\delta = 0.5, \gamma = 1$ to perform consistently well across datasets. All models were implemented in Python, and experiments were performed on an x64 machine with 2.4GHz Intel Xeon cores and 16 GB of memory. \vspace{-2pt}

\subsection{Prediction Tasks}
\label{sub:Prediction Tasks}
\begin{figure*}[!b]
 \centering
 \includegraphics[width=\textwidth]{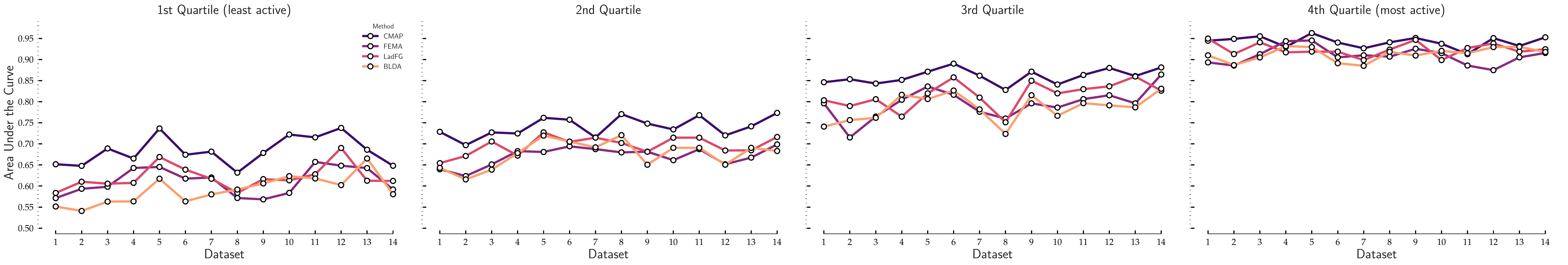}
 \vspace{-22pt}
 \caption{Effects of activity sparsity on prediction tasks (AUC) for Stack Exchanges (datasets 1-10) and MOOCs (datasets 11-14). CMAP has greatest performance gains in Quartile-1 (Sparse users), performance gap reduces for very active users (Quartile-4).}
 \label{data_sparsity}
\end{figure*}

In this section, we identify three prediction tasks, discuss evaluation metric and compare results with competing baseline methods. We focus on two platform specific user characterization tasks, and a common content-action prediction task.
\begin{description}
 \item \textbf{Certificate Prediction:} Students maintaining a minimum cumulative grade over course assignments are awarded certifications by Coursera. Connecting behavior to performance may help better online educational experiences. We attempt to predict students recieving certificates based on their behavioral data in each MOOC.
 \item \textbf{User Reputation Prediction:} For Stack-Exchanges, we predict if participants have a high reputation. Participants receive a reputation score based on the quality of their participation. We define high reputation as the top quartile of all reputation scores on that stack-exchange.
 \item \textbf{Behavior Prediction:} We predict the distribution of participant behavior across content and actions, in all Stack-Exchanges and MOOC datasets. Specifically, for each dataset, we
 extract $T=20$ topics from the text of all interactions using DMM~\cite{dmm}, and assign each participant interaction to the most likely topic learned by DMM.
\end{description}

We use standard classifiers and evaluation metrics to evaluate all models. Prediction tasks use linear kernel Support Vector Machine (SVM) classifier (with default parameters) in sklearn\footnote{http://scikit-learn.org/} and we compute results for each dataset through 10-fold cross validation.
\newpage
\begin{table}[!htbp]
 \vspace{-2pt}
 \begin{tabular}{@{}ccccc@{}}
  \toprule
  \textbf{Method} & \textbf{Precision}      & \textbf{Recall}         & \textbf{F1-score}       & \textbf{AUC}            \\
  \midrule
  LRC             & 0.76 $\pm$ 0.04          & 0.71 $\pm$ 0.05          & 0.74 $\pm$ 0.04          & 0.72 $\pm$ 0.03          \\
  DMM             & 0.77 $\pm$ 0.03          & 0.74 $\pm$ 0.04          & 0.75 $\pm$ 0.03          & 0.74 $\pm$ 0.03          \\
  LadFG           & 0.81 $\pm$ 0.02          & 0.78 $\pm$ 0.02          & 0.79 $\pm$ 0.02          & 0.79 $\pm$ 0.02          \\
  FEMA            & 0.78 $\pm$ 0.03          & 0.75 $\pm$ 0.04          & 0.76 $\pm$ 0.03          & 0.78 $\pm$ 0.03          \\
  CMAP            & \textbf{0.86 $\pm$ 0.02} & \textbf{0.81 $\pm$ 0.03} & \textbf{0.83 $\pm$ 0.02} & \textbf{0.84 $\pm$ 0.02} \\
  BLDA            & 0.80 $\pm$ 0.04          & 0.75 $\pm$ 0.03          & 0.77 $\pm$ 0.03          & 0.77 $\pm$ 0.04          \\
  \bottomrule
 \end{tabular}
 \vspace{1.5pt}
 \caption{Certification Prediction ($\mu \pm \sigma$ across MOOCs). CMAP outperforms baselines by 6.26-15.97\% AUC.}
 \label{tab:certificate prediction}
 \vspace{-29pt}
\end{table}

\begin{table}[!htbp]
 \centering
 \begin{tabular}{@{}ccccc@{}}
  \toprule
  \textbf{Method} & \textbf{Precision}       & \textbf{Recall}          & \textbf{F1-score}        & \textbf{AUC}             \\
  \midrule
  LRC             & 0.73 $\pm$ 0.04          & 0.69 $\pm$ 0.04          & 0.72 $\pm$ 0.03          & 0.73 $\pm$ 0.03          \\
  DMM             & 0.69 $\pm$ 0.05          & 0.65 $\pm$ 0.04          & 0.66 $\pm$ 0.04          & 0.70 $\pm$ 0.04          \\
  LadFG           & \textbf{0.86 $\pm$ 0.03} & 0.75 $\pm$ 0.03          & 0.79 $\pm$ 0.02          & 0.80 $\pm$ 0.03          \\
  FEMA            & 0.79 $\pm$ 0.04          & 0.73 $\pm$ 0.03          & 0.77$\pm$ 0.03           & 0.79 $\pm$ 0.04          \\
  CMAP            & 0.85 $\pm$ 0.02          & \textbf{0.83 $\pm$ 0.03} & \textbf{0.84 $\pm$ 0.02} & \textbf{0.86 $\pm$ 0.02} \\
  BLDA            & 0.75 $\pm$ 0.04          & 0.71 $\pm$ 0.04          & 0.74 $\pm$ 0.03          & 0.74 $\pm$ 0.04          \\
  \bottomrule
 \end{tabular}
 \vspace{1.5pt}
 \caption{Reputation Pred. ($\mu\pm\sigma$ across Stack-Exchanges). CMAP outperforms baselines by 6.65-21.43\% AUC. \vspace{7.5pt}}
 \label{tab:reputation prediction}
 \vspace{-29pt}
\end{table}
LRC is not used in behavior prediction since it does not build a user representation. We evaluated performance of all methods in the certificate and reputation prediction tasks via Precision, Recall, F1-Score and Area-Under-Curve (AUC). For the behavior prediction task, we measure the Root Mean Squared Error (RMSE) in the predicted user interaction proportions for $\textit{(topic, action)}$ pairs against the actual interaction proportions of users.

We show strong results across prediction tasks. In the certificate prediction task, our method improves on the baselines using the AUC measure by 6.26-15.97\%, averaged across MOOCs (c.f.~\Cref{tab:certificate prediction}). In the reputation prediction task over Stack-Exchanges, we improve on all baselines on the AUC metric in the range 6.65-21.43\%. In the behavior prediction task, our method improves on competing baselines using RMSE by 12\%-25\% on MOOCs and between 9.5\%-22\% on Stack-Exchange datasets (c.f.~\Cref{tab:behavior prediction}).

\subsection{Effects of Data Sparsity}
\label{sub:Effects of data sparsity}
In order to study the gains of our algorithm on characterizing users with different levels of activity, we split participants in each dataset into four equal partitions based on their number of interactions (Quartiles 1-4, 1 being least active). We then sample participants from each quartile and evaluate all methods on prediction performance (AUC).

Our model shows magnified gains (\Cref{data_sparsity}) in prediction performance over baseline models in the first and second quartiles which correspond to sparse or inactive users. BLDA performs the weakest in Quartile-1 since it relies on interaction activity to build user representations. Our model effectively bridges gaps in sparse or inconsistent participant data by exploiting similar active users within user partitions.

\begin{table}[t]
 \centering
 \begin{tabular}{@{}cccccc@{}}
  \toprule
  \textbf{Method} & \textbf{DMM}   & \textbf{LadFG} & \textbf{FEMA}  & \textbf{CMAP}           & \textbf{BLDA}  \\
  \midrule
  MOOC            & 4.9 $\pm$ 0.4  & 4.2 $\pm$0.3   & 4.1 $\pm$ 0.2  & \textbf{3.6  $\pm$ 0.2} & 4.4  $\pm$ 0.4 \\
  Stack-Ex        & 8.6  $\pm$ 0.6 & 7.9  $\pm$ 0.4 & 7.5  $\pm$ 0.3 & \textbf{6.7  $\pm$ 0.4} & 7.4  $\pm$ 0.5 \\
  \bottomrule
 \end{tabular}
 \vspace{1.5pt}
 \caption{Behavior Prediction (RMSE ($\times 10^{-2}$) $\mu \pm \sigma$). CMAP ouperforms baselines in MOOCs (12\%-25\%) and Stack-Exchanges (9.5\%-22\%)}
 \label{tab:behavior prediction}
 \vspace{-14pt}
\end{table}

\subsection{Effects of Behavior Skew}
\label{sub:Effects of behavioral skew}
We study the effect of behavioral skew on the prediction results, by subsampling users who predominantly perform the two most common activities in our two largest datasets, Ask-Ubuntu (Comments and Questions) and Comp Sci-1 (Play and Skip) in half, and retaining all other users. This reduces overall skew in the data. Baseline models are expected to perform better with de-skew. All models degrade in Ask-Ubuntu owing to significant content loss in the de-skew process.
\begin{table}[hbtp]
 \centering
 \vspace{-3.5pt}
 \begin{tabular}{@{}rcccc@{} }
  \toprule
  \multirow{2}{*}{\textbf{Method}} & \multicolumn{2}{@{}c}{\textbf{Ask-Ubuntu}} & \multicolumn{2}{c@{}}{\textbf{CompSci1 MOOC}}\\
  \cmidrule(lr){2-3} \cmidrule(lr){4-5}
        & Original        & Deskewed        & Original        & Deskewed        \\
  \midrule
  LRC   & 0.671          & 0.656          & 0.713          & 0.734          \\
  DMM   & 0.647          & 0.611          & 0.684          & 0.672          \\
  LadFG & 0.734         & 0.718          & 0.806          & 0.830          \\
  BLDA  & 0.706          & 0.683          & 0.739          & 0.788          \\
  CMAP  & \textbf{0.823} & \textbf{0.746} & \textbf{0.851} & \textbf{0.839} \\
  \bottomrule
 \end{tabular}
 \vspace{2pt}
 \caption{CMAP outperforms baselines (AUC) in original and de-skewed datasets. Performance gap reduces with de-skew.}
 \label{tab:skew}
 \vspace{-18pt}
\end{table}

We also investigate performance gains achieved by our approach in our most skewed and sparse Stack-Exchange (Ask-Ubuntu) vs least skewed (Christianity, \Cref{d1}). On average, our model outperforms baselines by 13.3\% AUC for Ask-Ubuntu vs 10.1\% for Christianity Stack-Exchange in the Reputation Prediction task.
\begin{table}[!htbp]
 \centering
 \vspace{-1pt}
 \begin{tabular}{@{}ccccccc@{}}
  \toprule
  \textbf{Method} & \textbf{DMM}  & \textbf{LRC} & \textbf{LadFG} & \textbf{FEMA}  & \textbf{CMAP}           & \textbf{BLDA}  \\
  \midrule
  Ask-Ubuntu            & 0.647 & 0.671  & 0.734   & -  & \textbf{0.823} & 0.706 \\
  Christianity        & 0.684 & 0.720 & 0.842 & 0.818 & \textbf{0.856} & 0.791 \\
  \bottomrule
 \end{tabular}
 \vspace{1.5pt}
 \caption{Performance gains in Reputation Pred for most skewed/sparse dataset (Ask-Ubuntu) vs least (Christianity)}
 \label{tab:behavior prediction}
 \vspace{-25pt}
\end{table}
%

\subsection{Parameter Sensitivity Analysis}
\label{sub:Parameter Sensitivity Analysis}
\begin{figure}[!htbp]
 \includegraphics[viewport = 1.25in 0in 16.75in 4.55in ,clip, width=\columnwidth]{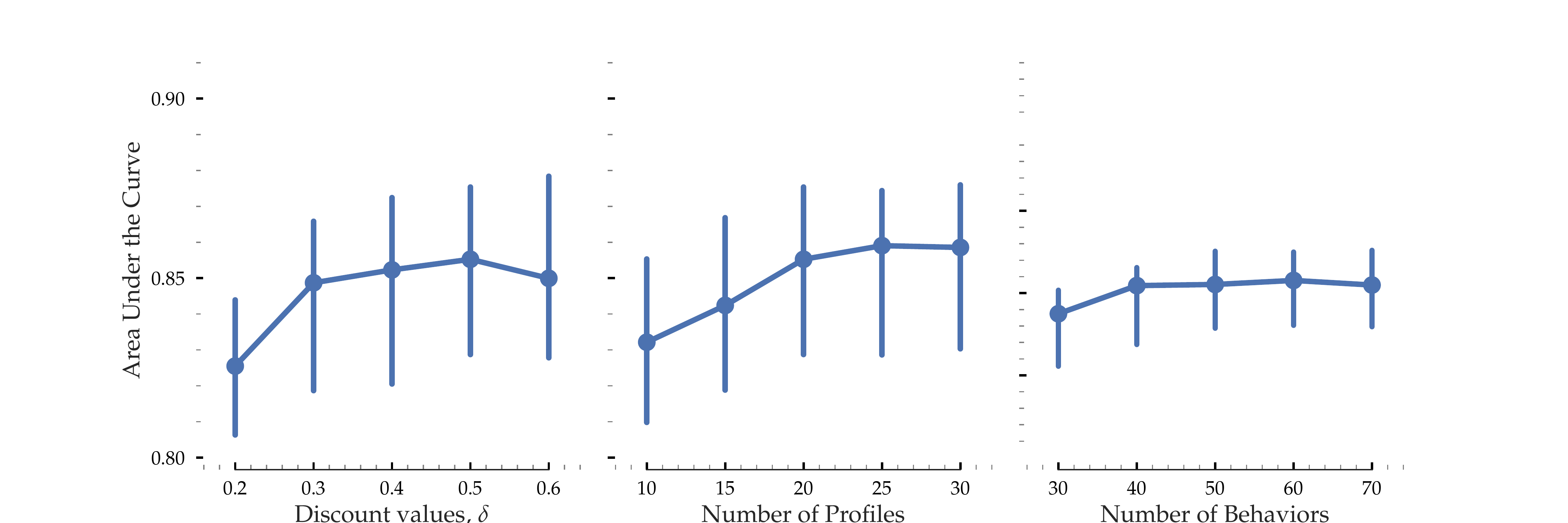}
 \caption{Mean performance(AUC) \& 95\% confidence interval with varying model parameters one at a time: $\delta$, $R$, $K$. Stability is observed in broad ranges of parameter values.}
 \label{fig:parameter sensitivity}
 \vspace{-10pt}
\end{figure}

Our model is primarily impacted by three parameter values: number of profiles $R$, number of behaviors $K$ and discount parameter $\delta$. We find results to be stable in a broad range of parameter values indicating that in practice our model requires minimal tuning (\Cref{fig:parameter sensitivity}). It is worth noting that while $R$ primarily impacts the granularity of the discovered activity profiles, while K impacts the resolution of content-action associations. Dirichlet parameters and other hyper-parameters have negligible impact on the profiles and behaviors learned. We set $R=20$, $\delta=0.5$ and $K=50$ for all datasets. Our inference algorithm is found to converge within 1\% AUC in less than 400 sampling iterations across all datasets.\vspace{-7pt}
\subsection{Scalability Analysis}
\label{sub:Scalability Analysis}
\begin{figure}[htbp]
 \includegraphics[width=\columnwidth]{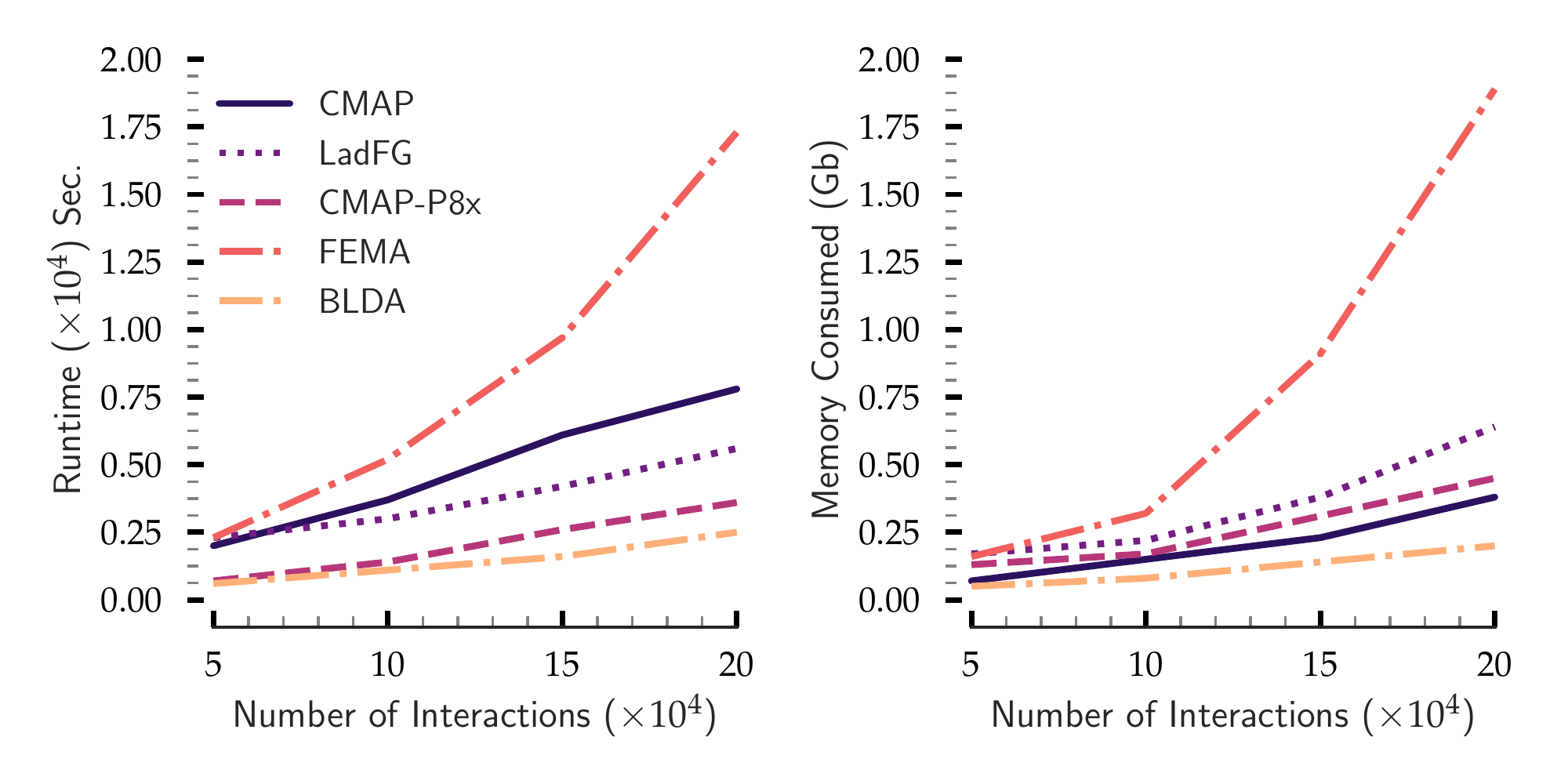}
 \vspace{-24pt}
 \caption{Effects of dataset size on algorithm runtime and memory consumption. BLDA is the fastest among the set of compared models. }
 \vspace{-5pt}
 \label{fig:memtime}
 \vspace{0pt}
\end{figure}

We compared the runtimes and memory consumption of our serial and batch-sampling (with 8 cores) inference algorithms with other models, for different volumes of interaction data obtained from random samples of the Ask-Ubuntu Stack-Exchange. BLDA is the fastest among the set of compared models. Our 8x batch sampler is comparable to BLDA in runtime. FEMA was the least scalable owing to the $|\mathcal{U}|^2$ growth of the User-User regularizer matrix. ~\Cref{fig:memtime} shows the comparisons between the algorithms.


\subsection{Limitations}
\label{limitations}
We identify two limitations. First, we make no assumptions about the structure of knowledge (e.g. a knowledge of ``probability'' is useful to understand ``statistical models''); incorporating knowledge structure, perhaps in the form of an appropriate prior will help with better understanding participants with low activity. Second, we assume a bounded time range and our model is inapplicable on streaming data.

\section{Related Work}
\label{sec:related work}
We categorize research related to our problem into four groups: Contextual Text Mining, Behavior Modeling, Temporal Behavior Dynamics, and Platform-specific work.\vspace{2.5pt}\\
\textbf{Contextual Text Mining}: Generative models which combine contextual information with text have found success in generating discriminative combined trends. Topics Over Time \cite{tot} is a latent generative model over text and time-stamps of documents. Other temporal content models have been proposed \cite{t1,t2}. Link based models attempt to extract a static view of author communities and content  \cite{tllda, l1,l2}. Short-text approaches \cite{dmm, satm} address content sparsity, absent publishing and consuming behaviors of users. Skew-aware models have also been developed for morphological structure analysis \cite{py1}, topic modeling \cite{pytm, pdlda, pap1}, dependency parsing \cite{py3} and query expansion \cite{dqe, zpap2}.  BLDA \cite{blda} is most closely related to our work since it attempts to integrate user actions with textual content, in the absence of a temporal factor. \vspace{1.5pt}\\
\textbf{Behavior Modeling}: Matrix factorization has been a popular approach to study and predict human behavior. In the past, models have attempted to study two dimensional relations such as user-item affiliation \cite{mf1,mf2}, and higher dimensional data via tensor models in web search and recommender systems \cite{tf1,tf2}. These models extract static views of user behavior.\\
\textbf{Temporal Behavior Dynamics}: This line of work integrates the temporal evolution of users with behavior modeling approaches. Previous works attempt to exploit historical user behavior data to predict future activity in online media \cite{bd1, bd2}, recommender applications \cite{bd3}, and academic communities \cite{bd4}. \cite{ladfg} attempts to build temporally-discretized latent representations of evolutionary learner behavior. These approaches do not explicitly address data sparsity at the user level. Jiang et al \cite{fema} have proposed a sparsity-aware tensor factorization approach to study user evolution. Their model however faces scalability challenges in massive real-world datasets, and relies on external regularizer data.\vspace{1.5pt}\\
\textbf{Platform-specific work}: Characterization of users with generated content and actions has been studied in both settings, MOOCs \cite{forum1, mooc2, mooc3} and community Question-Answering \cite{ stack2, trirole}. \cite{taxonomy} develops an engagement taxonomy for learner behavior. \cite{drop1, drop2} integrate social data to study dropout in MOOCs. More recently, adversarial learning has been applied to address the long-tail in Neural Collaborative Filtering \cite{cikm2018_short}. In contrast to these approaches, our objective is to learn robust and generalizable representations to study participant behavior in diverse interactive social learning platforms.

\section{Conclusion}
\label{sec:conclusion}
In this paper, we address the challenge of characterizing user behavior on social learning platforms in the presence of data sparsity and behavior skew. We proposed a CRP-based Multi-facet Activity Profiling model (CMAP) to profile user activity with both, content interactions as well as social ties. Our experimental results on diverse real-world datasets show that we outperform state of the art baselines on prediction tasks. We see strong gains on participants with low activity, our algorithms scale well, and perform well even with de-skewing data.

We identify three rewarding future directions. Developing Incremental models for streaming data; incorporating priors to structure knowledge; allowing for infinite action spaces.
%
%
%

\bibliographystyle{ACM-Reference-Format}
\bibliography{draft}


\begin{thebibliography}{00}


\ifx \showCODEN    \undefined \def \showCODEN     #1{\unskip}     \fi
\ifx \showDOI      \undefined \def \showDOI       #1{{\tt DOI:}\penalty0{#1}\ }
  \fi
\ifx \showISBNx    \undefined \def \showISBNx     #1{\unskip}     \fi
\ifx \showISBNxiii \undefined \def \showISBNxiii  #1{\unskip}     \fi
\ifx \showISSN     \undefined \def \showISSN      #1{\unskip}     \fi
\ifx \showLCCN     \undefined \def \showLCCN      #1{\unskip}     \fi
\ifx \shownote     \undefined \def \shownote      #1{#1}          \fi
\ifx \showarticletitle \undefined \def \showarticletitle #1{#1}   \fi
\ifx \showURL      \undefined \def \showURL       {\relax}        \fi
\providecommand\bibfield[2]{#2}
\providecommand\bibinfo[2]{#2}
\providecommand\natexlab[1]{#1}
\providecommand\showeprint[2][]{arXiv:#2}

\bibitem[\protect\citeauthoryear{Aldous, Ibragimov, and Jacod}{Aldous
  et~al\mbox{.}}{2006}]%
        {crp}
\bibfield{author}{\bibinfo{person}{David~J Aldous}, \bibinfo{person}{Illdar~A
  Ibragimov}, {and} \bibinfo{person}{Jean Jacod}.}
  \bibinfo{year}{2006}\natexlab{}.
\newblock \bibinfo{booktitle}{{\em Ecole d'Ete de Probabilites de Saint-Flour
  XIII, 1983}}. Vol.~\bibinfo{volume}{1117}.
\newblock \bibinfo{publisher}{Springer}.
\newblock


\bibitem[\protect\citeauthoryear{Anderson, Huttenlocher, Kleinberg, and
  Leskovec}{Anderson et~al\mbox{.}}{2014}]%
        {taxonomy}
\bibfield{author}{\bibinfo{person}{Ashton Anderson}, \bibinfo{person}{Daniel
  Huttenlocher}, \bibinfo{person}{Jon Kleinberg}, {and} \bibinfo{person}{Jure
  Leskovec}.} \bibinfo{year}{2014}\natexlab{}.
\newblock \showarticletitle{Engaging with massive online courses}. In
  \bibinfo{booktitle}{{\em Proceedings of the 23rd international conference on
  World wide web}}. ACM, \bibinfo{pages}{687--698}.
\newblock


\bibitem[\protect\citeauthoryear{Barab{\'a}si and Albert}{Barab{\'a}si and
  Albert}{1999}]%
        {Barabasi1999}
\bibfield{author}{\bibinfo{person}{Albert-L{\'a}szl{\'o} Barab{\'a}si} {and}
  \bibinfo{person}{R{\'e}ka Albert}.} \bibinfo{year}{1999}\natexlab{}.
\newblock \showarticletitle{Emergence of scaling in random networks}.
\newblock \bibinfo{journal}{{\em science\/}} \bibinfo{volume}{286},
  \bibinfo{number}{5439} (\bibinfo{year}{1999}), \bibinfo{pages}{509--512}.
\newblock


\bibitem[\protect\citeauthoryear{Bayer, Bydzovsk{\'a}, G{\'e}ryk, Obsivac, and
  Popelinsky}{Bayer et~al\mbox{.}}{2012}]%
        {drop1}
\bibfield{author}{\bibinfo{person}{Jaroslav Bayer}, \bibinfo{person}{Hana
  Bydzovsk{\'a}}, \bibinfo{person}{Jan G{\'e}ryk}, \bibinfo{person}{Tom{\'a}s
  Obsivac}, {and} \bibinfo{person}{Lubomir Popelinsky}.}
  \bibinfo{year}{2012}\natexlab{}.
\newblock \showarticletitle{Predicting Drop-Out from Social Behaviour of
  Students.}
\newblock \bibinfo{journal}{{\em International Educational Data Mining
  Society\/}} (\bibinfo{year}{2012}).
\newblock


\bibitem[\protect\citeauthoryear{Chang and Blei}{Chang and Blei}{2009}]%
        {l1}
\bibfield{author}{\bibinfo{person}{Jonathan Chang} {and}
  \bibinfo{person}{David~M Blei}.} \bibinfo{year}{2009}\natexlab{}.
\newblock \showarticletitle{Relational topic models for document networks}. In
  \bibinfo{booktitle}{{\em International conference on artificial intelligence
  and statistics}}. \bibinfo{pages}{81--88}.
\newblock


\bibitem[\protect\citeauthoryear{Coffrin, Corrin, de~Barba, and
  Kennedy}{Coffrin et~al\mbox{.}}{2014}]%
        {mooc3}
\bibfield{author}{\bibinfo{person}{Carleton Coffrin}, \bibinfo{person}{Linda
  Corrin}, \bibinfo{person}{Paula de Barba}, {and} \bibinfo{person}{Gregor
  Kennedy}.} \bibinfo{year}{2014}\natexlab{}.
\newblock \showarticletitle{Visualizing patterns of student engagement and
  performance in MOOCs}. In \bibinfo{booktitle}{{\em Proceedings of the fourth
  international conference on learning analytics and knowledge}}. ACM,
  \bibinfo{pages}{83--92}.
\newblock


\bibitem[\protect\citeauthoryear{Cui, Jin, Yu, Wang, Zhu, and Yang}{Cui
  et~al\mbox{.}}{2013}]%
        {bd1}
\bibfield{author}{\bibinfo{person}{Peng Cui}, \bibinfo{person}{Shifei Jin},
  \bibinfo{person}{Linyun Yu}, \bibinfo{person}{Fei Wang},
  \bibinfo{person}{Wenwu Zhu}, {and} \bibinfo{person}{Shiqiang Yang}.}
  \bibinfo{year}{2013}\natexlab{}.
\newblock \showarticletitle{Cascading outbreak prediction in networks: a
  data-driven approach}. In \bibinfo{booktitle}{{\em Proceedings of the 19th
  ACM SIGKDD international conference on Knowledge discovery and data mining}}.
  ACM, \bibinfo{pages}{901--909}.
\newblock


\bibitem[\protect\citeauthoryear{Cui, Wang, Liu, Ou, Yang, and Sun}{Cui
  et~al\mbox{.}}{2011}]%
        {bd3}
\bibfield{author}{\bibinfo{person}{Peng Cui}, \bibinfo{person}{Fei Wang},
  \bibinfo{person}{Shaowei Liu}, \bibinfo{person}{Mingdong Ou},
  \bibinfo{person}{Shiqiang Yang}, {and} \bibinfo{person}{Lifeng Sun}.}
  \bibinfo{year}{2011}\natexlab{}.
\newblock \showarticletitle{Who should share what?: item-level social influence
  prediction for users and posts ranking}. In \bibinfo{booktitle}{{\em
  Proceedings of the 34th international ACM SIGIR conference on Research and
  development in Information Retrieval}}. ACM, \bibinfo{pages}{185--194}.
\newblock


\bibitem[\protect\citeauthoryear{Diao, Jiang, Zhu, and Lim}{Diao
  et~al\mbox{.}}{2012}]%
        {dir2}
\bibfield{author}{\bibinfo{person}{Qiming Diao}, \bibinfo{person}{Jing Jiang},
  \bibinfo{person}{Feida Zhu}, {and} \bibinfo{person}{Ee-Peng Lim}.}
  \bibinfo{year}{2012}\natexlab{}.
\newblock \showarticletitle{Finding bursty topics from microblogs}. In
  \bibinfo{booktitle}{{\em Proceedings of the 50th Annual Meeting of the
  Association for Computational Linguistics: Long Papers-Volume 1}}.
  Association for Computational Linguistics, \bibinfo{pages}{536--544}.
\newblock


\bibitem[\protect\citeauthoryear{Goldwater, Johnson, and Griffiths}{Goldwater
  et~al\mbox{.}}{2006}]%
        {py1}
\bibfield{author}{\bibinfo{person}{Sharon Goldwater}, \bibinfo{person}{Mark
  Johnson}, {and} \bibinfo{person}{Thomas~L Griffiths}.}
  \bibinfo{year}{2006}\natexlab{}.
\newblock \showarticletitle{Interpolating between types and tokens by
  estimating power-law generators}. In \bibinfo{booktitle}{{\em Advances in
  neural information processing systems}}. \bibinfo{pages}{459--466}.
\newblock


\bibitem[\protect\citeauthoryear{Hu, Cao, Xu, Cao, Gu, and Zhu}{Hu
  et~al\mbox{.}}{2013}]%
        {tf1}
\bibfield{author}{\bibinfo{person}{Liang Hu}, \bibinfo{person}{Jian Cao},
  \bibinfo{person}{Guandong Xu}, \bibinfo{person}{Longbing Cao},
  \bibinfo{person}{Zhiping Gu}, {and} \bibinfo{person}{Can Zhu}.}
  \bibinfo{year}{2013}\natexlab{}.
\newblock \showarticletitle{Personalized recommendation via cross-domain
  triadic factorization}. In \bibinfo{booktitle}{{\em Proceedings of the 22nd
  international conference on World Wide Web}}. ACM, \bibinfo{pages}{595--606}.
\newblock


\bibitem[\protect\citeauthoryear{Ji, Xu, Wang, and He}{Ji
  et~al\mbox{.}}{2012}]%
        {stack2}
\bibfield{author}{\bibinfo{person}{Zongcheng Ji}, \bibinfo{person}{Fei Xu},
  \bibinfo{person}{Bin Wang}, {and} \bibinfo{person}{Ben He}.}
  \bibinfo{year}{2012}\natexlab{}.
\newblock \showarticletitle{Question-answer topic model for question retrieval
  in community question answering}. In \bibinfo{booktitle}{{\em Proceedings of
  the 21st ACM international conference on Information and knowledge
  management}}. ACM, \bibinfo{pages}{2471--2474}.
\newblock


\bibitem[\protect\citeauthoryear{Jiang, Cui, Wang, Xu, Zhu, and Yang}{Jiang
  et~al\mbox{.}}{2014}]%
        {fema}
\bibfield{author}{\bibinfo{person}{Meng Jiang}, \bibinfo{person}{Peng Cui},
  \bibinfo{person}{Fei Wang}, \bibinfo{person}{Xinran Xu},
  \bibinfo{person}{Wenwu Zhu}, {and} \bibinfo{person}{Shiqiang Yang}.}
  \bibinfo{year}{2014}\natexlab{}.
\newblock \showarticletitle{Fema: flexible evolutionary multi-faceted analysis
  for dynamic behavioral pattern discovery}. In \bibinfo{booktitle}{{\em
  Proceedings of the 20th ACM SIGKDD international conference on Knowledge
  discovery and data mining}}. ACM, \bibinfo{pages}{1186--1195}.
\newblock


\bibitem[\protect\citeauthoryear{Koren, Bell, and Volinsky}{Koren
  et~al\mbox{.}}{2009}]%
        {mf1}
\bibfield{author}{\bibinfo{person}{Yehuda Koren}, \bibinfo{person}{Robert
  Bell}, {and} \bibinfo{person}{Chris Volinsky}.}
  \bibinfo{year}{2009}\natexlab{}.
\newblock \showarticletitle{Matrix factorization techniques for recommender
  systems}.
\newblock \bibinfo{journal}{{\em Computer\/}} \bibinfo{volume}{42},
  \bibinfo{number}{8} (\bibinfo{year}{2009}).
\newblock


\bibitem[\protect\citeauthoryear{Krishnan, Deepak, Ranu, and Mehta}{Krishnan
  et~al\mbox{.}}{2017a}]%
        {zpap2}
\bibfield{author}{\bibinfo{person}{Adit Krishnan}, \bibinfo{person}{P Deepak},
  \bibinfo{person}{Sayan Ranu}, {and} \bibinfo{person}{Sameep Mehta}.}
  \bibinfo{year}{2017}\natexlab{a}.
\newblock \showarticletitle{Leveraging semantic resources in diversified query
  expansion}.
\newblock \bibinfo{journal}{{\em World Wide Web\/}} (\bibinfo{year}{2017}),
  \bibinfo{pages}{1--27}.
\newblock


\bibitem[\protect\citeauthoryear{Krishnan, Sankar, Zhi, and Han}{Krishnan
  et~al\mbox{.}}{2017b}]%
        {pap1}
\bibfield{author}{\bibinfo{person}{Adit Krishnan}, \bibinfo{person}{Aravind
  Sankar}, \bibinfo{person}{Shi Zhi}, {and} \bibinfo{person}{Jiawei Han}.}
  \bibinfo{year}{2017}\natexlab{b}.
\newblock \showarticletitle{Unsupervised Concept Categorization and Extraction
  from Scientific Document Titles}.
\newblock \bibinfo{journal}{{\em arXiv preprint arXiv:1710.02271\/}}
  (\bibinfo{year}{2017}).
\newblock


\bibitem[\protect\citeauthoryear{Krishnan, Sharma, Sankar, and
  Sundaram}{Krishnan et~al\mbox{.}}{2018b}]%
        {cikm2018_short}
\bibfield{author}{\bibinfo{person}{Adit Krishnan}, \bibinfo{person}{Ashish
  Sharma}, \bibinfo{person}{Aravind Sankar}, {and} \bibinfo{person}{Hari
  Sundaram}.} \bibinfo{year}{2018}\natexlab{b}.
\newblock \showarticletitle{An Adversarial Approach to Improve Long-Tail
  Performance in Neural Collaborative Filtering}. In \bibinfo{booktitle}{{\em
  Proceedings of the 27th ACM International Conference on Information and
  Knowledge Management}}. ACM, \bibinfo{pages}{1491--1494}.
\newblock


\bibitem[\protect\citeauthoryear{Krishnan, Sharma, and Sundaram}{Krishnan
  et~al\mbox{.}}{2018a}]%
        {cikm2018}
\bibfield{author}{\bibinfo{person}{Adit Krishnan}, \bibinfo{person}{Ashish
  Sharma}, {and} \bibinfo{person}{Hari Sundaram}.}
  \bibinfo{year}{2018}\natexlab{a}.
\newblock \showarticletitle{Insights from the Long-Tail: Learning Latent
  Representations of Online User Behavior in the Presence of Skew and
  Sparsity}. In \bibinfo{booktitle}{{\em Proceedings of the 27th ACM
  International Conference on Information and Knowledge Management}}. ACM,
  \bibinfo{pages}{297--306}.
\newblock


\bibitem[\protect\citeauthoryear{Leskovec, Huttenlocher, and
  Kleinberg}{Leskovec et~al\mbox{.}}{2010}]%
        {lrc}
\bibfield{author}{\bibinfo{person}{Jure Leskovec}, \bibinfo{person}{Daniel
  Huttenlocher}, {and} \bibinfo{person}{Jon Kleinberg}.}
  \bibinfo{year}{2010}\natexlab{}.
\newblock \showarticletitle{Predicting positive and negative links in online
  social networks}. In \bibinfo{booktitle}{{\em Proceedings of the 19th
  international conference on World wide web}}. ACM, \bibinfo{pages}{641--650}.
\newblock


\bibitem[\protect\citeauthoryear{Lindsey, Headden~III, and Stipicevic}{Lindsey
  et~al\mbox{.}}{2012}]%
        {pdlda}
\bibfield{author}{\bibinfo{person}{Robert~V Lindsey},
  \bibinfo{person}{William~P Headden~III}, {and} \bibinfo{person}{Michael~J
  Stipicevic}.} \bibinfo{year}{2012}\natexlab{}.
\newblock \showarticletitle{A phrase-discovering topic model using hierarchical
  pitman-yor processes}. In \bibinfo{booktitle}{{\em Proceedings of the 2012
  Joint Conference on Empirical Methods in Natural Language Processing and
  Computational Natural Language Learning}}. Association for Computational
  Linguistics, \bibinfo{pages}{214--222}.
\newblock


\bibitem[\protect\citeauthoryear{Liu}{Liu}{1994}]%
        {cgs}
\bibfield{author}{\bibinfo{person}{Jun~S Liu}.}
  \bibinfo{year}{1994}\natexlab{}.
\newblock \showarticletitle{The collapsed Gibbs sampler in Bayesian
  computations with applications to a gene regulation problem}.
\newblock \bibinfo{journal}{{\it J. Amer. Statist. Assoc.}}
  \bibinfo{volume}{89}, \bibinfo{number}{427} (\bibinfo{year}{1994}),
  \bibinfo{pages}{958--966}.
\newblock


\bibitem[\protect\citeauthoryear{Liu, Tang, Han, Jiang, and Yang}{Liu
  et~al\mbox{.}}{2010}]%
        {bd2}
\bibfield{author}{\bibinfo{person}{Lu Liu}, \bibinfo{person}{Jie Tang},
  \bibinfo{person}{Jiawei Han}, \bibinfo{person}{Meng Jiang}, {and}
  \bibinfo{person}{Shiqiang Yang}.} \bibinfo{year}{2010}\natexlab{}.
\newblock \showarticletitle{Mining topic-level influence in heterogeneous
  networks}. In \bibinfo{booktitle}{{\em Proceedings of the 19th ACM
  international conference on Information and knowledge management}}. ACM,
  \bibinfo{pages}{199--208}.
\newblock


\bibitem[\protect\citeauthoryear{Liu, Niculescu-Mizil, and Gryc}{Liu
  et~al\mbox{.}}{2009}]%
        {tllda}
\bibfield{author}{\bibinfo{person}{Yan Liu}, \bibinfo{person}{Alexandru
  Niculescu-Mizil}, {and} \bibinfo{person}{Wojciech Gryc}.}
  \bibinfo{year}{2009}\natexlab{}.
\newblock \showarticletitle{Topic-link LDA: joint models of topic and author
  community}. In \bibinfo{booktitle}{{\em proceedings of the 26th annual
  international conference on machine learning}}. ACM,
  \bibinfo{pages}{665--672}.
\newblock


\bibitem[\protect\citeauthoryear{Ma, Lyu, and King}{Ma et~al\mbox{.}}{2010}]%
        {dqe}
\bibfield{author}{\bibinfo{person}{Hao Ma}, \bibinfo{person}{Michael~R Lyu},
  {and} \bibinfo{person}{Irwin King}.} \bibinfo{year}{2010}\natexlab{}.
\newblock \showarticletitle{Diversifying Query Suggestion Results.}. In
  \bibinfo{booktitle}{{\em AAAI}}, Vol.~\bibinfo{volume}{10}.
\newblock


\bibitem[\protect\citeauthoryear{Ma, Sun, Yuan, and Cong}{Ma
  et~al\mbox{.}}{2015}]%
        {trirole}
\bibfield{author}{\bibinfo{person}{Zongyang Ma}, \bibinfo{person}{Aixin Sun},
  \bibinfo{person}{Quan Yuan}, {and} \bibinfo{person}{Gao Cong}.}
  \bibinfo{year}{2015}\natexlab{}.
\newblock \showarticletitle{A Tri-Role Topic Model for Domain-Specific Question
  Answering.}. In \bibinfo{booktitle}{{\em AAAI}}. \bibinfo{pages}{224--230}.
\newblock


\bibitem[\protect\citeauthoryear{Mackness, Mak, and Williams}{Mackness
  et~al\mbox{.}}{2010}]%
        {forum1}
\bibfield{author}{\bibinfo{person}{Jenny Mackness}, \bibinfo{person}{Sui Mak},
  {and} \bibinfo{person}{Roy Williams}.} \bibinfo{year}{2010}\natexlab{}.
\newblock \showarticletitle{The ideals and reality of participating in a MOOC}.
  In \bibinfo{booktitle}{{\em Proceedings of the 7th International Conference
  on Networked Learning 2010}}. University of Lancaster.
\newblock


\bibitem[\protect\citeauthoryear{Matsubara, Sakurai, Prakash, Li, and
  Faloutsos}{Matsubara et~al\mbox{.}}{2012}]%
        {t2}
\bibfield{author}{\bibinfo{person}{Yasuko Matsubara}, \bibinfo{person}{Yasushi
  Sakurai}, \bibinfo{person}{B~Aditya Prakash}, \bibinfo{person}{Lei Li}, {and}
  \bibinfo{person}{Christos Faloutsos}.} \bibinfo{year}{2012}\natexlab{}.
\newblock \showarticletitle{Rise and fall patterns of information diffusion:
  model and implications}. In \bibinfo{booktitle}{{\em Proceedings of the 18th
  ACM SIGKDD international conference on Knowledge discovery and data mining}}.
  ACM, \bibinfo{pages}{6--14}.
\newblock


\bibitem[\protect\citeauthoryear{Minka}{Minka}{2000}]%
        {fpi}
\bibfield{author}{\bibinfo{person}{Thomas Minka}.}
  \bibinfo{year}{2000}\natexlab{}.
\newblock \bibinfo{title}{Estimating a Dirichlet distribution}.
\newblock   (\bibinfo{year}{2000}).
\newblock


\bibitem[\protect\citeauthoryear{Pitman and Yor}{Pitman and Yor}{1997}]%
        {py}
\bibfield{author}{\bibinfo{person}{Jim Pitman} {and} \bibinfo{person}{Marc
  Yor}.} \bibinfo{year}{1997}\natexlab{}.
\newblock \showarticletitle{The two-parameter Poisson-Dirichlet distribution
  derived from a stable subordinator}.
\newblock \bibinfo{journal}{{\em The Annals of Probability\/}}
  (\bibinfo{year}{1997}), \bibinfo{pages}{855--900}.
\newblock


\bibitem[\protect\citeauthoryear{Qiu, Tang, Liu, Gong, Zhang, Zhang, and
  Xue}{Qiu et~al\mbox{.}}{2016}]%
        {ladfg}
\bibfield{author}{\bibinfo{person}{Jiezhong Qiu}, \bibinfo{person}{Jie Tang},
  \bibinfo{person}{Tracy~Xiao Liu}, \bibinfo{person}{Jie Gong},
  \bibinfo{person}{Chenhui Zhang}, \bibinfo{person}{Qian Zhang}, {and}
  \bibinfo{person}{Yufei Xue}.} \bibinfo{year}{2016}\natexlab{}.
\newblock \showarticletitle{Modeling and predicting learning behavior in
  MOOCs}. In \bibinfo{booktitle}{{\em Proceedings of the Ninth ACM
  International Conference on Web Search and Data Mining}}. ACM,
  \bibinfo{pages}{93--102}.
\newblock


\bibitem[\protect\citeauthoryear{Qiu, Zhu, and Jiang}{Qiu
  et~al\mbox{.}}{2013}]%
        {blda}
\bibfield{author}{\bibinfo{person}{Minghui Qiu}, \bibinfo{person}{Feida Zhu},
  {and} \bibinfo{person}{Jing Jiang}.} \bibinfo{year}{2013}\natexlab{}.
\newblock \showarticletitle{It is not just what we say, but how we say them:
  Lda-based behavior-topic model}. In \bibinfo{booktitle}{{\em Proceedings of
  the 2013 SIAM International Conference on Data Mining}}. SIAM,
  \bibinfo{pages}{794--802}.
\newblock


\bibitem[\protect\citeauthoryear{Qu, Chen, Jensen, and Skovsgaard}{Qu
  et~al\mbox{.}}{2015}]%
        {spacetime}
\bibfield{author}{\bibinfo{person}{Qiang Qu}, \bibinfo{person}{Cen Chen},
  \bibinfo{person}{Christian~S Jensen}, {and} \bibinfo{person}{Anders
  Skovsgaard}.} \bibinfo{year}{2015}\natexlab{}.
\newblock \showarticletitle{Space-Time Aware Behavioral Topic Modeling for
  Microblog Posts.}
\newblock \bibinfo{journal}{{\em IEEE Data Eng. Bull.\/}} \bibinfo{volume}{38},
  \bibinfo{number}{2} (\bibinfo{year}{2015}), \bibinfo{pages}{58--67}.
\newblock


\bibitem[\protect\citeauthoryear{Quan, Kit, Ge, and Pan}{Quan
  et~al\mbox{.}}{2015}]%
        {satm}
\bibfield{author}{\bibinfo{person}{Xiaojun Quan}, \bibinfo{person}{Chunyu Kit},
  \bibinfo{person}{Yong Ge}, {and} \bibinfo{person}{Sinno~Jialin Pan}.}
  \bibinfo{year}{2015}\natexlab{}.
\newblock \showarticletitle{Short and Sparse Text Topic Modeling via
  Self-Aggregation.}. In \bibinfo{booktitle}{{\em IJCAI}}.
  \bibinfo{pages}{2270--2276}.
\newblock


\bibitem[\protect\citeauthoryear{Ramesh, Goldwasser, Huang, Daum{\'e}~III, and
  Getoor}{Ramesh et~al\mbox{.}}{2013}]%
        {mooc2}
\bibfield{author}{\bibinfo{person}{Arti Ramesh}, \bibinfo{person}{Dan
  Goldwasser}, \bibinfo{person}{Bert Huang}, \bibinfo{person}{Hal
  Daum{\'e}~III}, {and} \bibinfo{person}{Lise Getoor}.}
  \bibinfo{year}{2013}\natexlab{}.
\newblock \showarticletitle{Modeling learner engagement in MOOCs using
  probabilistic soft logic}. In \bibinfo{booktitle}{{\em NIPS Workshop on Data
  Driven Education}}, Vol.~\bibinfo{volume}{21}. \bibinfo{pages}{62}.
\newblock


\bibitem[\protect\citeauthoryear{Riahi, Zolaktaf, Shafiei, and Milios}{Riahi
  et~al\mbox{.}}{2012}]%
        {qa}
\bibfield{author}{\bibinfo{person}{Fatemeh Riahi}, \bibinfo{person}{Zainab
  Zolaktaf}, \bibinfo{person}{Mahdi Shafiei}, {and} \bibinfo{person}{Evangelos
  Milios}.} \bibinfo{year}{2012}\natexlab{}.
\newblock \showarticletitle{Finding expert users in community question
  answering}. In \bibinfo{booktitle}{{\em Proceedings of the 21st International
  Conference on World Wide Web}}. ACM, \bibinfo{pages}{791--798}.
\newblock


\bibitem[\protect\citeauthoryear{Ros{\'e}, Carlson, Yang, Wen, Resnick,
  Goldman, and Sherer}{Ros{\'e} et~al\mbox{.}}{2014}]%
        {drop2}
\bibfield{author}{\bibinfo{person}{Carolyn~Penstein Ros{\'e}},
  \bibinfo{person}{Ryan Carlson}, \bibinfo{person}{Diyi Yang},
  \bibinfo{person}{Miaomiao Wen}, \bibinfo{person}{Lauren Resnick},
  \bibinfo{person}{Pam Goldman}, {and} \bibinfo{person}{Jennifer Sherer}.}
  \bibinfo{year}{2014}\natexlab{}.
\newblock \showarticletitle{Social factors that contribute to attrition in
  MOOCs}. In \bibinfo{booktitle}{{\em Proceedings of the first ACM conference
  on Learning@ scale conference}}. ACM, \bibinfo{pages}{197--198}.
\newblock


\bibitem[\protect\citeauthoryear{Ruan, Fuhry, and Parthasarathy}{Ruan
  et~al\mbox{.}}{2013}]%
        {l2}
\bibfield{author}{\bibinfo{person}{Yiye Ruan}, \bibinfo{person}{David Fuhry},
  {and} \bibinfo{person}{Srinivasan Parthasarathy}.}
  \bibinfo{year}{2013}\natexlab{}.
\newblock \showarticletitle{Efficient community detection in large networks
  using content and links}. In \bibinfo{booktitle}{{\em Proceedings of the 22nd
  international conference on World Wide Web}}. ACM,
  \bibinfo{pages}{1089--1098}.
\newblock


\bibitem[\protect\citeauthoryear{Saif, He, and Alani}{Saif
  et~al\mbox{.}}{2012}]%
        {sparse1}
\bibfield{author}{\bibinfo{person}{Hassan Saif}, \bibinfo{person}{Yulan He},
  {and} \bibinfo{person}{Harith Alani}.} \bibinfo{year}{2012}\natexlab{}.
\newblock \showarticletitle{Alleviating data sparsity for twitter sentiment
  analysis}. CEUR Workshop Proceedings (CEUR-WS. org).
\newblock


\bibitem[\protect\citeauthoryear{Sato and Nakagawa}{Sato and Nakagawa}{2010}]%
        {pytm}
\bibfield{author}{\bibinfo{person}{Issei Sato} {and} \bibinfo{person}{Hiroshi
  Nakagawa}.} \bibinfo{year}{2010}\natexlab{}.
\newblock \showarticletitle{Topic models with power-law using Pitman-Yor
  process}. In \bibinfo{booktitle}{{\em Proceedings of the 16th ACM SIGKDD
  international conference on Knowledge discovery and data mining}}. ACM,
  \bibinfo{pages}{673--682}.
\newblock


\bibitem[\protect\citeauthoryear{Sinha, Jermann, Li, and Dillenbourg}{Sinha
  et~al\mbox{.}}{2014}]%
        {mooc1}
\bibfield{author}{\bibinfo{person}{Tanmay Sinha}, \bibinfo{person}{Patrick
  Jermann}, \bibinfo{person}{Nan Li}, {and} \bibinfo{person}{Pierre
  Dillenbourg}.} \bibinfo{year}{2014}\natexlab{}.
\newblock \showarticletitle{Your click decides your fate: Inferring information
  processing and attrition behavior from MOOC video clickstream interactions}.
\newblock \bibinfo{journal}{{\em arXiv preprint arXiv:1407.7131\/}}
  (\bibinfo{year}{2014}).
\newblock


\bibitem[\protect\citeauthoryear{Sun, Zeng, Liu, Lu, and Chen}{Sun
  et~al\mbox{.}}{2005}]%
        {tf2}
\bibfield{author}{\bibinfo{person}{Jian-Tao Sun}, \bibinfo{person}{Hua-Jun
  Zeng}, \bibinfo{person}{Huan Liu}, \bibinfo{person}{Yuchang Lu}, {and}
  \bibinfo{person}{Zheng Chen}.} \bibinfo{year}{2005}\natexlab{}.
\newblock \showarticletitle{Cubesvd: a novel approach to personalized web
  search}. In \bibinfo{booktitle}{{\em Proceedings of the 14th international
  conference on World Wide Web}}. ACM, \bibinfo{pages}{382--390}.
\newblock


\bibitem[\protect\citeauthoryear{Teh}{Teh}{2006}]%
        {teh}
\bibfield{author}{\bibinfo{person}{Yee~Whye Teh}.}
  \bibinfo{year}{2006}\natexlab{}.
\newblock \showarticletitle{A hierarchical Bayesian language model based on
  Pitman-Yor processes}. In \bibinfo{booktitle}{{\em Proceedings of the 21st
  International Conference on Computational Linguistics and the 44th annual
  meeting of the Association for Computational Linguistics}}. Association for
  Computational Linguistics, \bibinfo{pages}{985--992}.
\newblock


\bibitem[\protect\citeauthoryear{Teh}{Teh}{2011}]%
        {dp}
\bibfield{author}{\bibinfo{person}{Yee~Whye Teh}.}
  \bibinfo{year}{2011}\natexlab{}.
\newblock \showarticletitle{Dirichlet process}.
\newblock In \bibinfo{booktitle}{{\em Encyclopedia of machine learning}}.
  \bibinfo{publisher}{Springer}, \bibinfo{pages}{280--287}.
\newblock


\bibitem[\protect\citeauthoryear{Wallach, Sutton, and McCallum}{Wallach
  et~al\mbox{.}}{2008}]%
        {py3}
\bibfield{author}{\bibinfo{person}{Hanna Wallach}, \bibinfo{person}{Charles
  Sutton}, {and} \bibinfo{person}{Andrew McCallum}.}
  \bibinfo{year}{2008}\natexlab{}.
\newblock \showarticletitle{Bayesian modeling of dependency trees using
  hierarchical Pitman-Yor priors}. In \bibinfo{booktitle}{{\em ICML Workshop on
  Prior Knowledge for Text and Language Processing}}. \bibinfo{pages}{15--20}.
\newblock


\bibitem[\protect\citeauthoryear{Wang and McCallum}{Wang and McCallum}{2006}]%
        {tot}
\bibfield{author}{\bibinfo{person}{Xuerui Wang} {and} \bibinfo{person}{Andrew
  McCallum}.} \bibinfo{year}{2006}\natexlab{}.
\newblock \showarticletitle{Topics over time: a non-Markov continuous-time
  model of topical trends}. In \bibinfo{booktitle}{{\em Proceedings of the 12th
  ACM SIGKDD international conference on Knowledge discovery and data mining}}.
  ACM, \bibinfo{pages}{424--433}.
\newblock


\bibitem[\protect\citeauthoryear{Wang, Zhai, and Roth}{Wang
  et~al\mbox{.}}{2013}]%
        {bd4}
\bibfield{author}{\bibinfo{person}{Xiaolong Wang}, \bibinfo{person}{Chengxiang
  Zhai}, {and} \bibinfo{person}{Dan Roth}.} \bibinfo{year}{2013}\natexlab{}.
\newblock \showarticletitle{Understanding evolution of research themes: a
  probabilistic generative model for citations}. In \bibinfo{booktitle}{{\em
  Proceedings of the 19th ACM SIGKDD international conference on Knowledge
  discovery and data mining}}. ACM, \bibinfo{pages}{1115--1123}.
\newblock


\bibitem[\protect\citeauthoryear{Yang and Leskovec}{Yang and Leskovec}{2011}]%
        {t1}
\bibfield{author}{\bibinfo{person}{Jaewon Yang} {and} \bibinfo{person}{Jure
  Leskovec}.} \bibinfo{year}{2011}\natexlab{}.
\newblock \showarticletitle{Patterns of temporal variation in online media}. In
  \bibinfo{booktitle}{{\em Proceedings of the fourth ACM international
  conference on Web search and data mining}}. ACM, \bibinfo{pages}{177--186}.
\newblock


\bibitem[\protect\citeauthoryear{Yin, Sun, Cui, Hu, and Chen}{Yin
  et~al\mbox{.}}{2013}]%
        {dir1}
\bibfield{author}{\bibinfo{person}{Hongzhi Yin}, \bibinfo{person}{Yizhou Sun},
  \bibinfo{person}{Bin Cui}, \bibinfo{person}{Zhiting Hu}, {and}
  \bibinfo{person}{Ling Chen}.} \bibinfo{year}{2013}\natexlab{}.
\newblock \showarticletitle{Lcars: a location-content-aware recommender
  system}. In \bibinfo{booktitle}{{\em Proceedings of the 19th ACM SIGKDD
  international conference on Knowledge discovery and data mining}}. ACM,
  \bibinfo{pages}{221--229}.
\newblock


\bibitem[\protect\citeauthoryear{Yin and Wang}{Yin and Wang}{2014}]%
        {dmm}
\bibfield{author}{\bibinfo{person}{Jianhua Yin} {and} \bibinfo{person}{Jianyong
  Wang}.} \bibinfo{year}{2014}\natexlab{}.
\newblock \showarticletitle{A dirichlet multinomial mixture model-based
  approach for short text clustering}. In \bibinfo{booktitle}{{\em Proceedings
  of the 20th ACM SIGKDD international conference on Knowledge discovery and
  data mining}}. ACM, \bibinfo{pages}{233--242}.
\newblock


\bibitem[\protect\citeauthoryear{Zhao, Cheng, Hong, and Chi}{Zhao
  et~al\mbox{.}}{2015}]%
        {www15}
\bibfield{author}{\bibinfo{person}{Zhe Zhao}, \bibinfo{person}{Zhiyuan Cheng},
  \bibinfo{person}{Lichan Hong}, {and} \bibinfo{person}{Ed~H Chi}.}
  \bibinfo{year}{2015}\natexlab{}.
\newblock \showarticletitle{Improving user topic interest profiles by behavior
  factorization}. In \bibinfo{booktitle}{{\em Proceedings of the 24th
  International Conference on World Wide Web}}. International World Wide Web
  Conferences Steering Committee, \bibinfo{pages}{1406--1416}.
\newblock


\bibitem[\protect\citeauthoryear{Zheng, Ding, Mamitsuka, and Zhu}{Zheng
  et~al\mbox{.}}{2013}]%
        {mf2}
\bibfield{author}{\bibinfo{person}{Xiaodong Zheng}, \bibinfo{person}{Hao Ding},
  \bibinfo{person}{Hiroshi Mamitsuka}, {and} \bibinfo{person}{Shanfeng Zhu}.}
  \bibinfo{year}{2013}\natexlab{}.
\newblock \showarticletitle{Collaborative matrix factorization with multiple
  similarities for predicting drug-target interactions}. In
  \bibinfo{booktitle}{{\em Proceedings of the 19th ACM SIGKDD international
  conference on Knowledge discovery and data mining}}. ACM,
  \bibinfo{pages}{1025--1033}.
\newblock


\end{thebibliography}

\end{document}